\documentclass[12pt]{article}
\usepackage{graphicx}
\usepackage{float}
\usepackage{xcolor}
\usepackage{tikz}
\usepackage[compat=1.1.0]{tikz-feynman}
\usepackage{framed}
\definecolor{shadecolor}{rgb}{0.90,0.90,0.90}
\usepackage{hyperref}
\usepackage{subcaption}
\usepackage[utf8]{inputenc}

\usepackage{setspace}
\usepackage{amsmath, amssymb, amsthm, float, graphicx}
\numberwithin{equation}{section}

\hypersetup{colorlinks=false, linkcolor=blue, citecolor=red}

\def\beq{\begin{eqnarray}}\def\eeq{\end{eqnarray}}
\def\be{\begin{equation}}\def\ee{\end{equation}}
\def\g{\gamma}

\def\m{\mu}

\def\a{\alpha}
\def\e{\epsilon}

\def\b{\beta}
\def\d{\delta}

\def\D{\Delta}
\def\G{\Gamma}
\def\l{\lambda}
\def\pd{\partial}

\def\bz{\bar{z}}

\def\mo{{\mathcal{O}}}

\def\mg{{\mathcal{G}}}
\def\mi{{\mathcal{I}}}

\def\G{\Gamma}

\usepackage{bm}
\usepackage{bm}

\begin{document}
\title{\bf Relating diagrammatic expansion with conformal correlator expansion}
\date{}

\author{Sunny $\text{Guha}^1$ and Kallol $\text{Sen}^{2,3}$\\~~~~\\
	${}^{1}$George P. and Cynthia Woods Mitchell Institute\\ for Fundamental Physics and Astronomy,\\
	Texas A\&M University, College Station, TX 77843, USA.
	\\~~~~\\
	${}^{2}$Kavli Institute for the Physics and Mathematics of the Universe (WPI),\\
	University of Tokyo, Kashiwa, Chiba 277-8583, Japan.
	\\~~\\
	${}^{3}$School of Mathematics, Trinity College Dublin, Dublin 2, Ireland.}

\maketitle
\vskip 2cm
\abstract{In this note we study the possible connection between functions appearing in diagrammatic expansion and the conformal correlator expansion. To study the connection we propose a generating function which can be expanded to construct a basis. This basis can be utilized to expand,  I) the four point function of scalars near the Wilson-Fisher fixed point in $d=4-\e$ as in \cite{Alday:2017zzv} and II) integrals for loop diagrams for massless $\phi^4$ theory in position space in four dimensions. This suggests that a linear combination of one expansion can be recast in terms of a linear combination of the other. As a by-product, we also derive the Mellin space representation for the twist-2 higher spin conformal blocks. We also discuss the higher derivative contact terms in the present scenario.}

\vfill {\footnotesize 
	sunnyguha@physics.tamu.edu  \ \ kallolmax@gmail.com}

\newpage

\newpage

\tableofcontents

\onehalfspacing
\section{Introduction}

The $\e$-expansion technique of Wilson and Kogut \cite{wilson1, wilson2} and innumerable followups demonstrated an effective way to compute the corrections to dynamical quantities (dimensions, coupling etc.) along the RG flow. Recently, \cite{Rychkov:2015naa,Sen:2015doa} developed tools to compute the same quantities from the bootstrap program. Equivalently, using the inversion formula of \cite{Caron-Huot:2017vep} in the light-cone limit $(1-\bz\ll z\ll 1)$,
\be\label{lsp}
C(\b)=\int d^2 z\ \m(z,\bz)\  k_\b(\bz)\ {\rm dDisc} \left[\mathcal{G}^t(z,\bz)\right]_{\text{LC}}+(t\rightarrow u)\,,
\ee
\cite{Alday:2017zzv, Henriksson:2018myn, Alday:2019clp} demonstrated that the tower of large spin double twist operators in the direct channel ($\b=\tau_\ell+2\ell$), is controlled by the perturbative expansion of the crossed channel correlator around a Wilson-Fisher fixed point in $d=4-\e$ dimensions\footnote{See also \cite{Alday:2016jfr, Alday:2016njk, Cardona:2018dov, Cardona:2018qrt} for useful applications related to large spin expansion and inversion formula.}. The quantity of specific interest is ${\rm dDisc} \left(\mathcal{G}^t(z,\bz)\right)$ expanded around the WF fixed point,
\be
{\rm dDisc}\left[\mathcal{G}^t(z,\bz)\right]_\text{LC}={\rm dDisc} \left[\frac{(z\bz)^{\D_\phi}}{(1-\bz)^{\D_\phi}}\sum_\mo C_{\phi\phi\mo}G_{\D_\mo,\ell}(1-z,1-\bz)\right]_\text{LC}\,,
\ee
where $\mo$ is the traceless symmetric exchange in the OPE of $\phi \times \phi$ around the perturbative fixed point and $G_{\D_\mo,\ell}$ is the conformal block. We have the following perturbative expansion for the scaling dimensions and three-point coefficients, 
\begin{align}\label{pertexp}
\begin{split}
\Delta_{\phi}&=\Delta^0_{\phi}+g^2 \g_{\phi}+\dots\,,\\
\Delta_\mo&= 2\Delta_{\phi}+\ell+\begin{cases}g \g_\mo\,, \ell=0\\ g^2\g_\mo\,, \ell>0\end{cases}+\dots\,,\\
C_{\phi \phi \mo} &= C^0_{\phi \phi \mo}+gC^1_{\phi \phi \mo} + \dots\,,
\end{split}    
\end{align}
where we treat $g$ and $\e$ to be independent for now. We represent the double-discontinuity as, 
\be\label{pertexp1}
{\rm dDisc}\left[\mathcal{G}^t(z,\bz)\right]_\text{LC}={\rm dDisc}\left[(1-\bz)^{\g/2}\right]\sum_{i+j=n} g^i\e^j f_{i,j}(z,\bz)\,,\ n\geq0\,,
\ee 
where $\g$ is the anomalous dimension given in \eqref{pertexp}. This is either $O(g)$ or $O(g^2)$ depending on scalar and higher spin exchanges respectively. Due to the perturbative expansion leading ($\tau=2$) scalar contributes to lowest orders upto $O(g^3)$ while $O(g^4)$ onwards scalars mix with other higher spin operators.  

In this note, we aim to understand whether (and to what extent) there exists a connection between the basis of functions $f_{i,j}(z,\bz)$ and the functions appearing in diagrammatic perturbation theory of massless $\phi^4$ theories. To sum up, we want to recast the expansion of ${\rm dDisc}\, \mg^t(z,\bz)$ in terms of functions inspired from the integrals found in diagrammatic massless $\phi^4$ theory.  We discuss the main results of the work below.\\

\vspace{0.5cm}
\textbf{Main results:}\\

\vspace{0.1cm}
There are three families in the conformal correlator expansion,\footnote{Our basis differs from the one in \cite{Alday:2017zzv}, since we want to use the entire function $f_{i,j}$ as a basis element}, 
\begin{align}
\begin{split}
O(g^{2+n}):&\  \{f_{n,0}(z,\bz)\}\,,\ O(g^2\e^n):\ \{f_{0,n}(z,\bz)\}\,,\ O(g^{2+i}\e^j):\ \{f_{i,j}(z,\bz)\}\,.
\end{split}
\end{align}
We constructed a basis which determines the minimal number of functions required for the conformal correlator expansion.  There exits a ``{\it generating function}" $\mathfrak{I}_{g_1,g_2,g_3}(z,\bz)$ which when expanded in $\delta$ gives rise to the basis mentioned above. $\delta$ is a small expansion parameter of the generating function (discussed in the main text). The basis are characterized by their transcendality, which is equal to their order of expansion in $\delta$. 
\be
I^n_{(g_1,g_2,g_3)}(z,\bz)=\pd_\d^n\mathfrak{I}_{g_1,g_2,g_3}(z,\bz)\bigg|_{\d=0}\,, 
\ee
where $z\bz=(x_{12}^2x_{34}^2)/(x_{13}^2x_{24}^2)\,, (1-z)(1-\bz)=(x_{14}^2x_{23}^2)/(x_{13}^2x_{24}^2)$. At each order in $n$, we can write a vector of functions characterized by the set $(g_1,g_2,g_3)$. In this basis we can write the expansion of the scalar four point function upto $O(g^4)$ as,
\begin{align}\label{mainresult}
\begin{split}
&O(g^2): -\log(1-\bar{z})^2\frac{B_0}{4}\,, \ \  O(g^3): \log(1-\bar{z})^2 \frac{B_1}{4}-\log(1-\bar{z})^3\frac{B_0}{24}\,, \\
&O(g^4): \frac{\log(1-\bar{z})^2}{48}(2B_1-B_2+(22+6\zeta_2-\log z\bz) B_0) +\log(1-\bar{z})^3\frac{B_1}{24}-\log(1-\bar{z})^4\frac{B_0}{192}\,, \\
&O(g^3\epsilon) : \log^2(1-\bar{z})\Big(\frac{H_2}{48}+(\log z \bar{z}+4)\frac{B_1}{16}-(\zeta_2-1) \frac{B_0}{4}\Big)+\frac{\log^3(1-\bar{z})}{6}\Big(\frac{B_0}{4}+\frac{\zeta_2}{4}+\log(z)\frac{B_0}{8}\Big)\,,\\
&O(g^2\epsilon): \log^2(1-\bar{z})\Big(\frac{B_0}{4}+\frac{\zeta_2}{4}+\log(z)\frac{B_0}{8}\Big)\,,  \\
&O(g^2\epsilon^2): -\log^2(1-\bar{z})\Big(\left(\frac{B_0 \zeta _2}{16}+\frac{\zeta _2}{4}-\frac{\zeta _3}{8}\right)+\left(\frac{B_0}{8}+\frac{\zeta _2}{8}\right) \log (z)+\frac{1}{32} B_0 \log
^2(z)\Big)\,.
\end{split}
\end{align}
At $O(g^4)$ we also have twist two contribution,
\begin{equation}
O(g^4): \log ^{2}(1-\overline{z})\left( \frac{B_2}{6}-(2+\zeta_2)B_0- \frac{1}{2} \log(z \bar{z})B_1  \right)\,.
\end{equation}
The $O(g^5)$ results are messy and lengthier and we avoid them here. On the other side we can also cast the finite pieces of loop integral in terms of the same basis,
\begin{align}
\begin{split}
&\text{tree}=\log(1-\bar{z})\frac{B_0}{4}\,,\ \ \text{1-loop}=\log^2(1-\bar{z})\left(\frac{B_0}{4}\right)+\log(1-\bar{z})\left( B_0-\frac{B_1}{2} \right)\,,\\
\text{2-loop}=& \log^3(1-\bar{z})\left( -\frac{B_0}{24}\right)+\log^2(1-\bar{z})\left(\frac{B_1}{4} \right) - \log(1-\bar{z})\left( \frac{B_2}{12}+(1-\frac{\zeta_2}{2})B_0 \right)\,.
\end{split}
\end{align}

The basis $B_n$ and $H_n$ is defined in the main next. Our main observation is that similar functions appear at $O(L+2)$ in conformal correlator expansion and at $L$-loop of perturbative diagram expansion.

Rest of the paper is organized as follows: In section \ref{expblock} and \ref{twist2} we discuss the expansion of the scalar and the twist$-2$ blocks to obtain the functions $f_{i,j}$. Section \ref{genf} discusses the the generating function relevant for the basis computation (specific structures appearing at each order in the expansion). Section \ref{dpt} discusses the matching between the functions of section \ref{genf} and \ref{expblock} and \ref{twist2} for upto $O(g^4)$. We comment on the structural properties of the functions appearing at the next higher order. In section \ref{contactt} we comment on how the contact interactions can be incorporated in this picture and the possible implications. We end the work with discussions and further questions  in section{discussion}. Appendix \ref{impid} contains relevant integrals. \ref{cbdetails} contain the details of the conformal blocks. Appendix \ref{fddetails} contain explicit details on the perturbative diagrams for constructing the generating function upto $3-$loops with an emphasis on the regularization scheme employed in \ref{regp}. Appendix \ref{li2} contains further input in the $3-$loop case.

\section{Scalar Block Expansion}\label{expblock}
To expand the conformal block we first start with the integral representation of the block with scalar exchange\cite{Dolan:2000ut, Dolan:2011dv, Dolan:2003hv}. We expand it in the $z\rightarrow 0$ limit and obtain (see -\ref{cbdetails} for details),
\begin{align}\label{scalb}
\begin{split}
&G_{\D,0}(1-z,1-\bz)\\
&=\frac{(1-\bz)^{\D/2}}{\bz}\frac{\G(1+\D-h)\G(\D)}{\G(\frac{\D}{2})^3\G(\frac{\D}{2}-h+2)}\bigg[(\D/2-(h-1-\D/2)\bz)\\
&\times\int_0^1 dx\ \mathcal{I}_1^{\D,h}(x,0,1-\bz)\bigg(2H_{\D/2-1}+\log\frac{z\bz}{(1-x(1-\bz))^2}\bigg)\\
&+(1-\bz)\int_0^1 dx\ \mathcal{I}_2^{\D,h}(x,0,1-\bz)\bigg(2+(h-2-\D/2)(2H_{\D/2-1}+\log\frac{z\bz}{(1-x(1-\bz))^2})\bigg)\bigg] \,.
\end{split}
\end{align}
In this expansion we plug in the parameters,
\be\label{def0}
d=4-\e\,, \ \D_\phi=(d-2)/2+\g_{\phi}(g)\,, \ \D=2\D_\phi+\g_\D(g)=d-2+\g(g)\,, 
\ee
where $\g=2\g_\phi+\g_\D$ and $h=d/2$. $\g_\D$ and $\g_\phi$ are respectively, 
\be\label{anomdim}
\g_\D(g)=g\,, \ \ \g_\phi(g)=\frac{g^2}{12}-\frac{g^3}{8}+\frac{11}{144}g^2\e+\dots\,.
\ee
Even though $g=f(\e)$, \footnote{Obtained by setting the $\b-$function to zero order by order in usual perturbative QFT.}for now we will consider these two parameters as being independent. Using \eqref{def0}, we expand \eqref{scalb} to get,
\begin{align}\label{scalarge}
\begin{split}
&\mathcal{G}_{\D,0}(1-z,1-\bz)=\bigg(\frac{z\bz}{1-\bz}\bigg)^{\D_\phi}G_{\D,0}(1-z,1-\bz)\\
&=(z\bz)^{\D_\phi}\frac{(1-\bz)^{g/2}}{\bz}\frac{\G(2-\e+\g(g))\G(1-\e/2+\g(g))}{\G(1+\frac{\g(g)-\e}{2})^3\G(1+\g(g)/2)}\bigg[(1-\bz)\int_0^1 dx \mathcal{I}_2^{\g(g),\e}(x,1-\bz)\\
&\hspace{5cm}+(1+(\g(g)-\e)/2+\g(g)/2\ \bz)\int_0^1 dx \mathcal{I}_1^{\g(g),\e}(x,1-\bz)\bigg]\,. 
\end{split}
\end{align}
where, 
\begin{align}
\begin{split}
\mathcal{I}_1^{\g(g),\e}(x,1-\bz)&=\frac{x^{\frac{\g(g)-\e}{2}}(1-x)^{\g(g)/2}}{(1-x(1-\bz))^{1+\frac{\g(g)-\e}{2}}}\bigg[2H_{\frac{\g(g)-\e}{2}}+\log\frac{z\bz}{(1-x(1-\bz))^2}\bigg]\,,\\
\mathcal{I}_2^{\g(g),\e}(x,1-\bz)&=\frac{x^{1+\frac{\g(g)-\e}{2}}(1-x)^{\g(g)/2}}{(1-x(1-\bz))^{1+\frac{\g(g)-\e}{2}}}\bigg[2-(1+\g(g)/2)\bigg(2H_{\frac{\g(g)-\e}{2}}+\log\frac{z\bz}{(1-x(1-\bz))^2}\bigg)\bigg]\,.
\end{split}
\end{align}
We utilize the following identity of harmonic numbers,
\be
H_{\frac{\g(g)-\e}{2}}=\sum_{n=1}^\infty (-1)^{n-1}\zeta_{n+1}\bigg(\frac{\g(g)-\e}{2}\bigg)^n\,,
\ee
and,
\be
\frac{x^{\frac{\g(g)-\e}{2}}(1-x)^{\g(g)/2}}{(1-x(1-\bz))^{\frac{\g(g)-\e}{2}}}=\sum_{m,n=0}^\infty \frac{(\frac{\g(g)}{2})^m (\frac{\e}{2})^n}{m!n!}\log^m\frac{x(1-x)}{1-x(1-\bz)}\log^n\frac{1-x(1-\bz)}{x}\,.
\ee
Using the expansion of anomalous dimension \cite{Dey:2016mcs}, 
\be
\g(g)=g+\a_1 g^2+\a_2 g^3+\a_3 g^2\e+\dots\,, \ \ \a_1=\frac{1}{6}\,, \ \a_2=-\frac{1}{4}\,, \ \a_3=\frac{11}{72}\,,
\ee
we rewrite \eqref{scalarge} in terms of the known integrals \cite{hypexp}, 
\be
I_{\chi_1}(\chi_2,\chi_3,\chi_4,1-\bz)=\int_0^1 \frac{x^{\chi_1}\log^{\chi_2}x\log^{\chi_3}(1-x)\log^{\chi_4}(1-x(1-\bz))}{1-x(1-\bz)}\,,
\ee
where, for our purposes $\chi_1=0,1$. Finally we obtain,
\be\label{ddiscscal}
{\rm dDisc}[G_{\D,0}(1-z,1-\bz)]=(z\bar{z})^{\Delta_{\phi}^{cl}}\bigg(\sum_{n=2}^\infty \frac{g^n}{2^n n!}\log^n(1-\bz)\bigg) \sum_{\a=0}^\infty\sum_{i+j=\a} g^i\e^j f_{i,j}(z,\bz)\,. 
\ee
The ${\rm dDisc}$ starts from $n\geq2$. The details of $f_{i,j}(z,\bz)$ are given in appendix \ref{basisel} and we intend to put these functions in a basis. In the following section, we perform the same analysis for twist$-2$ higher spin ($\ell\geq 2$) operators.

\section{Twist$-2$ higher spin operators}\label{twist2}
In the $z\rightarrow0$ limit, the twist-2 block is given by (refer to \eqref{ft4} of \ref{cbdetails}),  
\begin{align}
\begin{split}
\mathcal{G}_{2}(z,\bz)=&(z\bz)^{\D_\phi}[(1-\bz)(1-z)]^{\l_2-\D_\phi}\int_0^1 dx \frac{(x(1-x))^{d/2-2}}{(z+x(1-z))^{d/2-1}(\bz+x(1-\bz))^{d/2-1}}\\
&\times \frac{\G(d-3)}{\G(d/2-1)^2}\sum_{\ell=2}^\infty \frac{(d-3+2\ell) }{\ell^2(\ell+1)^2}C_\ell^{(d-3)/2}(1-2x)\,.
\end{split}
\end{align}
For the order we are interested in, $\G(d-3)/\G(d/2-1)^2=1$. For twist$-2$ operators the scaling dimensions are $\D=2\D_\phi+\ell+g^2 \g_\ell$ and hence $\l_2=(\D-\ell)/2=\D_\phi+g^2/2 \g_\ell$ and $\D_\phi=1-\e/2$. To perform the sum over spin we consider a generalization of the form,
\be
\sum_{\ell=2}^\infty \frac{(2\l+2\ell)}{(J_\l^2)^m}C_\ell^\l(x)=F_{m,\l}(x)\,,
\ee
where for twist-2 operators $J_\l^2=(\ell+\l+1/2)(\ell+\l-1/2)$, $\l=(d-3)/2$ and $d=4-\e$. Using the differential equation \cite{Alday:2016njk} for $C^\l_\ell(x)$, 
\be
D_\l\equiv (1-x^2)d_x^2-(2\l+1)x d_x-(\l^2-1/4)\,,\  D_\l C^\l_\ell(x)=-J_\l^2C^\l_\ell(x)\,, \ d_x\equiv d/dx\,,
\ee
we can write,
\be\label{gendiff}
D_\l F_{m,\l}(x)=-F_{m-1,\l}(x)\,,\ \ \forall \ m\geq1\,,
\ee
as the generalization of the leading contribution. First we determine $F_{0,\l}(x)$ and obtain the boundary conditions for $F_{1,\l}$ and $F_{2,\l}$ and plug the lower order solutions in the {\it rhs} to determine the $m+1$-th terms. Regarding the boundary conditions one can show that,
\be
F'_{m,\l}(x=0)=0\,.
\ee
Using the integral representation of $C_\ell^\l(x)$, 
\begin{align}\label{init}
\begin{split}
C_\ell^\l(x)&=\oint dz(1-2x z+z^2)^{-\l} z^{-1-\ell}\,, \\
F_{0,\l}(x)&= \sum_{\ell=2}^\infty (2\l+2\ell) C^\l_\ell(x)=-2\l\,.
\end{split}
\end{align}
Solving for the first order we find, 
\be
F_{1,\l}(x)=a_{0,0}-\frac{1}{2}\log(1-x^2)+\frac{\e}{8}(8a_{0,1}+2(a_{0,0}+1)\log(1-x^2)-\log^2(1-x)-\log^2(1+x))\,,
\ee
where,
\be
F_{1,\l}(0)=a_{0,0}+\e\ a_{0,1}\,, 
\ee
and the constrants $a_{0,0}$ and $a_{0,1}$ can be determined from solving \eqref{init} at $x=0$ and expanding in $\lambda=(1-\e)/2$. For $m=2$ a direct evaluation as a function of $\l$ can be challenging, hence we separate, 
\be\label{f2lambda}
F_{2,\l}(x)=g(x)+\e\ h(x)\,,
\ee
where, 
\begin{align}
\begin{split}
g(x)=&b_{0,0}+\frac{1+a_{0,0}}{2}\log(1-x^2)-\frac{1}{2}\log(1-x)\log(1+x)\,,\\
h(x)=&\frac{1}{24}\bigg[6\tanh^{-1}x\bigg(\zeta_2-2{\rm Li}_2\frac{x-1}{x+1}\bigg)-3\log^2(1-x)\log(1+x)-\log^3(1+x)\\
&-12{\rm Li}_3\frac{x-1}{x+1}+3\log^2(1-x^2)(1+a_{0,0})+6\log(1-x^2)(2a_{0,1}+b_{0,0})+24 b_{0,1}-9\zeta_3\bigg]\,.
\end{split}
\end{align}
We also have,
\be
F_{2,\l}(0)=b_{0,0}+\e\ b_{0,1}\,.
\ee
Here $b_{0,0}$ and $b_{0,1}$ are constants to be detemined by plugging in \eqref{gendiff} in \eqref{init} and expanding at $x=0$ around $\l=(1-\e)/2$. We will use the representation,
\be
\tanh^{-1}x=\frac{1}{2}\log\frac{1+x}{1-x}\,.
\ee
In terms of these decompostions, we can write \eqref{ft4} in the form (putting $d=4-\e$),
\begin{align}\label{twistexp}
\begin{split}
\mathcal{G}_{2}(z,\bz)=&(z\bz)^{\D_\phi}((1-\bz)(1-z))^{g^2/2}\int_0^1 \frac{dx}{(z+x(1-z))(\bz+x(1-\bz))}\\
&\times \bigg(1+\frac{\e}{2}\log\frac{(z+x(1-z))(\bz+x(1-\bz))}{x(1-x)}\bigg)[g(1-2x)+\e h(1-2x)]\,,\\
=&(z\bz)^{\D^\text{cl}_\phi}((1-\bz)(1-z))^{g^2/2}(F_0(z,\bz)+\e F_1(z,\bz))\,.
\end{split}
\end{align}
The individual parts are,
\begin{align}
\begin{split}
F_0(z,\bz)=&\int_0^1 dx\frac{g(1-2x)}{(z+x(1-z))(\bz+x(1-\bz))}\,,\\
F_1(z,\bz)=&\int_0^1 \frac{dx}{(z+x(1-z))(\bz+x(1-\bz))}\bigg(h(1-2x)+\frac{g(1-2x)}{2}\log\frac{(z+x(1-z))(\bz+x(1-\bz))}{z\bz[x(1-x)]}\bigg)\,.
\end{split}
\end{align}
We see from (\ref{twistexp}) that the double discontinuities start from $O(g^4)$. $F_0$ appears at $O(g^4)$ and $F_1$ appears at $O(g^4\e)$. We can put $F_1$ in the following format, 
\begin{align}
\begin{split}
&h(1-2x)+\frac{g(1-2x)}{2}\log\frac{(z+x(1-z))(\bz+x(1-\bz))}{z\bz[x(1-x)]}\\
=&\frac{1}{24}\bigg[C_0+3(C_1+2\zeta_2-4)\log x(1-x)+3\zeta_2\log\frac{1-x}{x}+6(2-\zeta_2)\log((z+x(1-z))(\bz+x(1-\bz)))\\
&-6\log x\log(1-x)\log((z+x(1-z))(\bz+x(1-\bz)))-\log^3(1-x)-3\log(1-x)\log^2x \\
&+6\log x\log(1-x)\log x(1-x)-6\bigg(\log\frac{1-x}{x}{\rm Li}_2\bigg(\frac{x}{x-1}\bigg)+2{\rm Li}_3\bigg(\frac{x}{x-1}\bigg)\bigg)\bigg]-\frac{\log z\bz}{2}g(1-2x)\,,
\end{split}
\end{align}
and the constants, 
\begin{align}
\begin{split}
C_0&=24b_{0,1}+24a_{0,1}\log 2 +6\log 2(2-\zeta_2)+2\log^3 2-9\zeta_3\,,\\
C_1&=4a_{0,1}-\zeta_2+2+\log^22\,.
\end{split}
\end{align}
Most of the integrals above can be split into a general form as discussed in appendix \ref{integrals}, where we provide a list of such integrals. For the $O(1)$ contribution, we know $a_{0,0}=\log2-1$ and $b_{0,0}=1-\zeta_2/2-\log^22/2$, so that $g(x)$ in \eqref{f2lambda} becomes,
\be
g(1-2x)=1-\frac{\zeta_2}{2}-\frac{\log x\log(1-x)}{2}\,.
\ee
In terms of the integrals we thus have,
\begin{align}
\begin{split}
F_0(z,\bz)&=\bigg(1-\frac{\zeta_2}{2}\bigg)\mathcal{I}_{0,0,0}(z,\bz)-\frac{1}{2}\mathcal{I}_{1,1,0}(z,\bz)\,,
\end{split}
\end{align}
while for $O(g^4\e)$, we can write, 
\begin{align}
\begin{split}
F_1(z,\bz)=&\frac{1}{24}(C_0\mathcal{I}_{0,0,0}+3(C_1+2\zeta_2-4)(\mathcal{I}_{1,0,0}+\mathcal{I}_{0,1,0})+3\zeta_2(\mathcal{I}_{0,1,0}-\mathcal{I}_{1,0,0})-\mathcal{I}_{0,3,0}\\
&+3\mathcal{I}_{2,1,0}+6\mathcal{I}_{1,2,0}+6((2-\zeta_2)\mathcal{J}_1+\mathcal{J}_2-\mathcal{J}_3))-\frac{\log z\bz}{2}F_0(z,\bz)\,.
\end{split}
\end{align}
We will eventually take the $z\rightarrow0$ limit so that, 
\be\label{mellint2}
\lim_{z\rightarrow0}F_0(z,\bz)=-\frac{1}{2\bz}\bigg((\zeta_2-2)\log z+2\log\bz+\frac{1}{6}\log^3\bz+{\rm Li}_3(1-\bz)-{\rm Li}_3\bigg(\frac{\bz-1}{\bz}\bigg)-\zeta_3\bigg)\,.
\ee
which matches with B.1 of \cite{Alday:2017zzv} modulo overall factors. For the first sub-leading term (including the first order expansion of $\bz^{\D_\phi}F_0(z,\bz)$), 
\begin{align}\label{mellint3}
\begin{split}
\lim_{z\rightarrow0}F_1(z,\bz)=&\frac{1}{24\bz}\bigg[3(8-\log^2\bz){\rm Li}_2(1-\bz)-3{\rm Li}_2(1-\bz)^2+6\log\bz{\rm Li}_3\bigg(\frac{\bz-1}{\bz}\bigg)\\
&+3(\zeta_2-2)(\log^2z+4\log z\log\bz-6\log^2\bz)-6\log^2\bz-12\zeta_2{\rm Li}_2(1-\bz)\\
&-\log^4\bz+9\zeta_4-6S_{2,2}(1-\bz)\bigg]+C_0\frac{\log\bz-\log z}{24\bz}\\
&+\frac{C_1+2\zeta_2-4}{4\bz}(\zeta_2-{\rm Li}_2(1-\bz))+(4-\zeta_2-C_1)\frac{\log^2\bz-\log^2 z}{16\bz}\,.
\end{split}
\end{align}
The above two equations are the main results of this section. (\ref{mellint2}) is the twist-2 contribution at $O(g^4)$ and (\ref{mellint3}) is the contribution of twist-2 at $O(g^4\epsilon)$.

\section{Generating Function}\label{genf}
A large portion (if not all) of the functional basis for the conformal block expansion can be generated by a ``{\it generating function}" of the form,
\begin{equation}\label{genreg}
\mathfrak{I}_{(g_1,g_2,g_3)}(z,\bz)={\rm Disc}\left[\int \frac{d^4x_6}{ x_{16}^{2+2g_1\delta} \, x_{26}^{2+2g_2\delta} \, x_{36}^{2+2g_3\delta} \, x_{46}^{2-2(g_1+g_2+g_3)\delta}}\right]_{z\rightarrow 0}\,,
\end{equation}
where $z\bz=(x_{12}^2x_{34}^2)/(x_{13}^2x_{24}^2)\,, (1-z)(1-\bz)=(x_{14}^2x_{23}^2)/(x_{13}^2x_{24}^2)$. We expand the generating function in $\d$ (using HypExp MATHEMATICA package) and at each order we can consider, 
\be
I^n_{(g_1,g_2,g_3)}(z,\bz)= \pd_\d^n \mathfrak{I}_{g_1,g_2,g_3}(z,\bz)\bigg|_{\d=0}\,.
\ee
For example, $I^0_{0,0,0}=\log z-\log\bz=B_0$ which is the basis function at the zeroth order. As we will demonstrate, both the diagrammatic perturbation and the conformal correlator expansion can be written in terms of the functions derived from $\mathfrak{I}_{(g_1,g_2,g_3)}$. The set $\{g_i\}$, provides considerable freedom for construction. However for our purposes,
\be
\text{I})\ g_1=g_3\ \&\ g_1=-g_2\,, \ \text{II})\ g_1=-g_2 \ \&\ g_3=(2-\sqrt{3})g_1\,.
\ee
covers most of the expense. For $n\geq1$, we denote the two classes as\footnote{We would like to stress that while the choice is not unique, it suffices our purpose.},
\begin{equation}
I^n_{(1,-1,1)}=B_n \qquad I^n_{(1,-1,2-\sqrt{3})}=H_n\,.
\end{equation}
The evaluation of the integral (\ref{genreg}) is listed in appendix-(\ref{masterintegsec}). We will write the results of this till first order,
\begin{align}
\begin{split}
=&\log(1-\bar{z})\left(\frac{\log(z)-\log(\bar{z})}{\bar{z}}\right)\, \\
&\Big(\log(1-\bar{z})( \frac{\text{Li}_2(1-\bar{z})-\zeta_2}{ \bar{z}} ) + \log(1-\bar{z})^2(\frac{\log (\bar{z})-\log (z)}{2\bar{z}})\Big) \delta\,.
\end{split}
\end{align}
At each $\delta^n$, the power of $\log(1-\bar{z})$ goes from unity to $n+1$.  However at each $n$, the new addition to the basis comes from the functional coefficient accompanying $\log(1-\bz)$. All the higher powers of $\log(1-\bz)$ are accompanied by functions which already appeared at lower order in $n$. Thus, we will construct our basis from the lowest order discontinuity. For class I), we find to few lowest orders,
\begin{align}\label{bbasis}
\begin{split}
&B_0=\log(z)-\log(\bar{z})\,, B_1 = \text{Li}_2(1-\bar{z})-\zeta_2\,, \\
& B_2 = -6 \zeta _3+6 \text{Li}_3(1-\bar{z})-6 \text{Li}_3\left(\frac{\bar{z}-1}{\bar{z}}\right)+3 \text{Li}_2(1-\bar{z}) \log
(z\bar{z})+9 \zeta _2(\log(z)- \log (\bar{z}))+\log ^3(\bar{z})\,, \\
&B_3 = 21 \zeta _4-24 \text{Li}_4(1-\bar{z})-6 \text{Li}_3(1-\bar{z})(\log(z) - \log (\bar{z}))+12 S_{2,2}(1-\bar{z})+6 \zeta _3(\log(z)- \log
(\bar{z})) \\
& \qquad -12\zeta_2(\text{Li}_2(1-\bar{z})-\zeta_2)\,,
\end{split}
\end{align}
while for class II),
\begin{align}\label{hbasis}
\begin{split}
&H_0=B_0\,, H_1=B_1\,,\\
&H_2=12 \zeta _3+3 \text{Li}_2(1-\bar{z}) \log (z\bz)-12 \text{Li}_3(1-\bar{z})-6
\text{Li}_3\left(\frac{\bar{z}-1}{\bar{z}}\right)+3 \zeta _2 (\log (z)-\log (\bar{z}))+\log ^3(\bar{z})\,.
\end{split}
\end{align}

Now we will try to argue why the generating function (\ref{genfunc}) seems a plausible choice.

\subsection{Connection to loop integrals}
The generating function we advocate is inspired by the class of integrals used to represent loop diagrams \cite{ud1,ud2,Drummond:2006rz} A particular class of integrals can be used to represent a large subset of loop diagrams (rings, sunsets etc. see appendix \ref{fddetails})\footnote{There are other class of integrals for ladder diagrams and convolutions of these integrals therein.}. This class of integrals are,
\begin{align}\label{genfunc}
\begin{split}
&\text{even-loop: } I_L=\lim_{\delta\rightarrow 0}(\frac{x_{23}^2}{x_{14}^2})^{\delta/2} f(\delta)^{L/2}f(-\delta)^{L/2}\int \frac{d^4x_6}{ x_{16}^{2-\delta} \, x_{46}^{2-\delta} \, x_{26}^{2+\delta} \, x_{36}^{2+\delta}}\,,\\
&\text{odd-loop: } I_L=\lim_{\delta\rightarrow 0}(\frac{x_{23}^2}{x_{14}^2})^{\delta/2} f(\delta)^{(L+1)/2}f(-\delta)^{(L-1)/2}\int \frac{d^4x_6}{ x_{16}^{2-\delta} \, x_{46}^{2-\delta} \, x_{26}^{2+\delta} \, x_{36}^{2+\delta}}\,,\\
\end{split}
\end{align}
where,
\begin{equation}
f(\delta)=\frac{\Gamma(\delta)\Gamma(1-\frac{\delta}{2})^2}{\Gamma(1+\frac{\delta}{2})^2\Gamma(2-\delta)}\,.
\end{equation}
Apart from the pre-factors, the final integral that needs to be done is the same. For $\d\rightarrow0$, the finite piece is obtained by expanding the integral in $\d$ which cancels the poles (in $\d$) coming from the prefactor.

As a check of the generating function we can rewrite the loop integrals in appendix \ref{fddetails} using the basis above. For tree level and one-loop we find,
\be\label{loopres1}
\text{tree}=\log(1-\bar{z})\frac{B_0}{4}\,,\ \ \text{1-loop}=\log^2(1-\bar{z})\left(\frac{B_0}{4}\right)+\log(1-\bar{z})\left( B_0-\frac{B_1}{2} \right)\,.
\ee
For two-loop ring,
\begin{align}\label{loopres2}
\begin{split}
\text{2-loop}=& \log^3(1-\bar{z})\left( -\frac{B_0}{24}\right)+\log^2(1-\bar{z})\left(\frac{B_1}{4} \right) - \log(1-\bar{z})\left( \frac{B_2}{12}+(1-\frac{\zeta_2}{2})B_0 \right)\,.
\end{split}
\end{align}
To conclude the section, note that \eqref{genreg} is a generalization of the class of integrals in \eqref{genfunc}, where we extended the notion of expansion to multiple parameters $\{g_i\}$ to allow for considerable freedom to construct a basis. The correspondence between loop expansion and  conformal correlator expansion suggests that the dDisc at $O(L+2)$ from the correlator expansion associates with Disc at $O(L)-$ diagrams. For example, the leading dDisc at $O(g^2)$ term of the CFT correlator associates with $O(g)$ term in the tree level, $O(g^3)$ connects with $1-$loop and so on. 

\section{Conformal correlator expansion}\label{dpt}
In this section we show that the conformal correlator expansion (\ref{ddiscscal}) can be cast in terms of the basis obtained in the previous section. We will split the contributions into three types - pure $g$ terms, $g^2\epsilon^n$ terms and everything else. Pure $g$ terms corresponds to expansion at fixed $d=4$ and we find that these can be obtained from ring-diagrams evaluated at $d=4$. In the comparisons we will always ignore the overall $\bar{z}$ factor. In particular for all order-4 terms ($O(g^4)$,$O(g^3\epsilon)$ and $O(g^2\epsilon^2)$) in the conformal correlator expansion the generating function (\ref{genfunc}) should be sufficient as we will see in the next section. We further refine our statement by saying that all the basis can be generated by very small number of generating functions for any given order. In fact we see that till $O(4)$ \eqref{genreg} suffices while an additional generating function is required at the next order\footnote{The additional generating function is associated with the ladder diagrams.}. 

\subsection{Pure $g$ pieces}

We first list down the pure $g$ contributions of the correlator,

\begin{align}
\begin{split}
&O(g^2): -\log(1-\bar{z})^2\left(\frac{\log (z)-\log (\bar{z})}{4 \bar{z}}\right) \,, \\
&O(g^3): \log(1-\bar{z})^2 \frac{\text{Li}_2(1-\bar{z})-\zeta _2}{4 \bar{z}}-\log(1-\bar{z})^3\bigg( \frac{\log (z)-\log (\bar{z})}{24 \bar{z}} \bigg) \,, \\
O(g^4): &\log(1-\bar{z})^2\bigg(\frac{6 \zeta _3-6 \text{Li}_3(1-\bar{z})+6 \text{Li}_3\left(\frac{\bar{z}-1}{\bar{z}}\right)-3 \text{Li}_2(1-\bar{z})
	\log (z \bar{z})+3 \zeta _2 (\log (\bar{z})-\log(z))-\log ^3(\bar{z})}{48 \bar{z}} \\
& +\frac{- \zeta _2+ \text{Li}_2(1-\bar{z})}{24 \bar{z}} -\frac{1}{48 \bar{z}}\log(z \bar{z})(\log(z)-\log(\bar{z})) + \frac{22 \log (z)-22 \log (\bar{z})}{48 \bar{z}}\bigg)\\
&+\log(1-\bar{z})^3(\frac{\text{Li}_2(1-\bar{z})-\zeta _2}{24 \bar{z}})-\log(1-\bar{z})^4(\frac{\log (z)-\log (\bar{z})}{192 \bar{z}})\,.
\end{split}
\end{align}

With the correlator expansion in hand we can now cast then in the basis constructed in (\ref{bbasis}),
\begin{align}
\begin{split}
&O(g^2): -\log(1-\bar{z})^2\frac{B_0}{4}\,, \ \ O(g^3): \log(1-\bar{z})^2 \frac{B_1}{4}-\log(1-\bar{z})^3\frac{B_0}{24} \,,\\
&O(g^4): \log(1-\bar{z})^2\bigg( -\frac{1}{48}B_2+\frac{1}{24}B_1+\frac{(11+3\zeta_2)}{24}B_0-\frac{1}{48}\log(z \bar{z})B_0\bigg)\\
&\qquad \qquad +\log(1-\bar{z})^3\frac{B_1}{24}-\log(1-\bar{z})^4\frac{B_0}{192}\,, \\
&O(g^5): \log(1-\bar{z})^2\bigg(\frac{3}{288}B_3-\frac{1}{144}B_2-\frac{(25+6\zeta_2)}{48}B_1+(-\frac{1}{16}-\frac{3}{4}\zeta_3+\frac{1}{24}\zeta_2)B_0+\log(z\bar{z})(\frac{1}{48}B_1 +\frac{1}{32}B_0) \bigg)\\
&\qquad \qquad+\log(1-\bar{z})^3\bigg( -\frac{1}{288}B_2+\frac{1}{144}B_1+\frac{(11+3\zeta_2)}{144}B_0-\frac{1}{288}\log(z \bar{z})B_0\bigg)\\
&\qquad \qquad+\log(1-\bar{z})^4\frac{B_1}{192}-\log(1-\bar{z})^5\frac{B_0}{1920}\,.
\end{split}
\end{align}

We notice that in all the comparisons we were able to write the conformal block expansion completely in terms of $B$-terms. The coefficients were just constants or $\zeta$ functions and the $\log(z \bar{z})$ term is an artifact of the kinematic factor.
We also observe an interesting pattern, the discontinuities at a given order $g^n$ appear as higher-discontinuities at order $g^{n+1}$ upto overall coefficients. The new information at every order is always contained in its lowest discontinuity (or the coefficient of $\log(1-\bar{z})^2$). The results obtained above are similar to loop diagram results in (\ref{loopres1},\ref{loopres2}). More specifically we notice that functions that appear at loop-$L$ are the same one that appear in conformal correlator at $O(L+2)$. 
\subsubsection{Prediction for O($g^6$)}
With the observation noted in the previous section we can make a prediction for next order in $g$,
\begin{align}
\begin{split}
O(g^6)&: \epsilon \log(1-\bar{z})^6B_0+\delta \log(1-\bar{z})^5B_1+\gamma \log(1-\bar{z})^4\bigg( B_1-\frac{1}{2}B_2+\frac{22+6\zeta_2-\log z \bar{z}}{2}B_0\bigg)\\ 
& +\beta \log(1-\bar{z})^3\bigg( \frac{B_3}{96}-\frac{B_2}{144}-\frac{(25+6\zeta_2-\log z\bz)}{48}B_1+\frac{4\zeta_2-72\zeta_3+3\log z\bz-6}{96}B_0\bigg)\\
&+\alpha \log(1-\bar{z})^2\bigg( B_4 \cdots  \bigg)\,,
\end{split}
\end{align}
where ($\alpha,\beta,\gamma,\delta,\epsilon $) are unfixed numerical coefficients. Only the lowest order discontinuity is unknown, however we do know that it is composed of a combination of $B_4$ and lower order $B_i$s. 
\subsubsection{Twist-2 Matching}
We will also try to put the twist-2 contribution (\ref{mellint2}) at O($g^4$) in terms of our basis. The O($g^4$) contribution is,
\begin{equation}
T_{2}= \log ^{2}(1-\overline{z})\left(\log z\left(\zeta_{2}-2\right)+2 \log \overline{z}+\frac{1}{6} \log ^{3} \overline{z}+\operatorname{Li}_{3}(1-\overline{z})-\operatorname{Li}_{3}\left(\frac{\overline{z}-1}{\overline{z}}\right)-\zeta_{3}\right)\,.
\end{equation}
In terms of our basis we can cast this as,
\begin{equation}
T_2=\left( \frac{B_2}{6}-(2+\zeta_2)B_0- \frac{1}{2} \log(z \bar{z})B_1  \right)\log ^{2}(1-\overline{z})\,.
\end{equation}
\subsection{$O(g^2\epsilon^n)$}
Here we will report an interesting observation regarding terms of type $g^2\epsilon^n$. Since these terms have $g^2$ they only contain a $\log^2(1-\bar{z})$ discontinuity.
In our perturbative diagram computations we have worked in $d=4$ dimensions instead of $d=4-\epsilon$. Working in $d=4-\epsilon$ would have given us $\epsilon$ corrections to our basis and we believe that this would be the honest way to generate a basis for $O(g^2\epsilon^n)$ terms. However, we can still get away with it because we notice nice pattern in the terms of this type. Since these contributions are simple enough we can cast them in their own basis,
\begin{align}
\begin{split}
&O(g^2\epsilon): \log^2(1-\bar{z})\left(\frac{B_0}{4}+\frac{\zeta_2}{4}+\log(z)\frac{B_0}{8}\right)\,,  \\
&O(g^2\epsilon^2): \log^2(1-\bar{z})\left(\left(\frac{B_0 \zeta _2}{16}+\frac{\zeta _2}{4}-\frac{\zeta _3}{8}\right)+\left(\frac{B_0}{8}+\frac{\zeta _2}{8}\right) \log (z)+\frac{1}{32} B_0 \log
^2(z)\right)\,, \\
O(g^2\epsilon^3): &\log^2(1-\bar{z})\Big[-\frac{\zeta _3}{8}+\frac{7 \zeta _4}{32}-B_0 \left(\frac{\zeta _3}{16}-\frac{\zeta _2}{16}\right)-\left(\frac{B_0 \zeta _2}{32}+\frac{1}{2} \left(\frac{\zeta
	_2}{4}-\frac{\zeta _3}{8}\right)\right) \log (z) \\
& - \frac{1}{192} B_0 \log ^3(z)+\frac{1}{32} \left(B_0+\zeta _2\right) \log ^2(z)\Big]\,.
\end{split}
\end{align}
We noticed a similar pattern appearing as the pure $g$ terms. At each order $g^2\epsilon^n$ there are terms with increasing power of $\log(z)$ which becomes terms of higher powers of $\log(z)$ in the next order. We make a final comment that the basis comprises of terms of form,
\begin{equation}
\equiv \{ B_0,\log^n(z)\} +  \{\zeta(n+1),\cdots , \zeta(2)\}|_{n\geq1}\,.
\end{equation}
\subsection{Remaining Terms}
 At fourth order there are 3 possible contributions to the conformal block expansion - $O(g^4),O(g^3\epsilon)$ and $O(g^2\epsilon^2)$. We have already cast $O(g^4)$ and $O(g^2\epsilon^2)$ in a basis and are left with $O(g^3\epsilon)$ term whose contribution is,
\begin{align}
\begin{split}
& O(g^3\epsilon) :  \log^3(1-\bar{z})\left(\frac{2 \zeta _2+(\log (z)+2) (\log (z)-\log (\bar{z}))}{48 \bar{z}} \right) \\
&-\frac{\log^2(1-\bar{z})}{48 \bar{z}} \bigg(12 \zeta _3+6 \text{Li}_2(1-\bar{z}) \log (z)+12 \text{Li}_2(1-\bar{z})-12 \text{Li}_3(1-\bar{z})-6
\text{Li}_3\left(\frac{\bar{z}-1}{\bar{z}}\right)\\
& +6 \text{Li}_2(1-\bar{z}) \log (\bar{z})-6 \zeta _2 (2 \log (z)-\log
(\bar{z})+2)+12 \log (z)+\log ^3(\bar{z})-12 \log (\bar{z}) \bigg)\,.
\end{split}
\end{align}
At this point we remind ourself of the second choice of regularization which resulted in an additional basis $H_2$ (\ref{hbasis}). We have already encountered the $\log^3(1-\bar{z})$ piece before. So here we will focus on only the $\log^2(1-\bar{z})$ term. It can be cast into a basis as follows,
\begin{equation}
O(g^3\epsilon)|_{\log^2(1-\bar{z})} : \frac{1}{48}H_2+\log(z \bar{z})\frac{B_1}{16}-\zeta_2 \frac{B_0}{4}+\frac{B_1+B_0}{4} \,.
\end{equation}
With this result in hand we have been able to show our basis covers the expansion till $O(4)$. This implies that our one generating function (\ref{genfunc}) is sufficient for all terms upto $O(4)$. At the next order we have terms $O(g^5)$, $O(g^4\epsilon)$, $O(g^3\epsilon^2)$ and $O(g^2\epsilon^3)$. While the first and last of the above term already fit in our basis,$O(g^4\epsilon)$ and $O(g^3\epsilon^2)$  (in \ref{listofcontrib}) has additional $\text{Li}_2^2(1-\bar{z})$ contribution. A similar issue arises for the $O(g^4\e)$ piece of twist$-2$ block as well. Note that the correspondence we have drawn, suggests that $O(5)$ in conformal block expansion should correspond to $3-$loop diagrams. At $3-$loop level there exists an additional generating function (from ladder diagrams) which accommodates $O(g^4\epsilon)$ and $O(g^3\epsilon^2)$. We have not performed the computation explicitly, but schematically show in Appendix(\ref{li2}), how the ladder diagram contributes a factor of $\text{Li}_2^2(1-\bar{z})$. 

\section{Incorporating the contact terms}\label{contactt}
We end by discussing the effect of the contact terms in the above picture. Following \cite{Gopakumar:2018xqi} the general representation of the contact terms are,
\be
A_c(z,\bz)=v^{-\D_\phi} \int ds dt v^s u^t \G(-t)^2\G(s+t)^2\G(\D_\phi-s)^2 M_c(s,t)\,.
\ee
where the overall $v^{-\D_\phi}$ factor is required to match with our loop results.
\be
M_c(s,t)=\sum_{\{m,n\}=0}^{m+n=L/2} a_{mn} [s(s+t-\D_\phi)(t+\D_\phi)]^m [t(s+t)+s(s-\D_\phi)]^n\,.
\ee
The first contact term is $\phi^4$ for which $\D_\phi=1-\e/2$ and $M_c(s,t)=a_{00}$ a constant. This integral we have already evaluated which gives,
\be
A_c(z,\bz)=\frac{\G(\D_\phi)^4}{\G(2\D_\phi)} \int_0^1 \frac{dx}{1-x(1-v)}\left(\frac{x(1-x)}{1-x(1-v)}\right)^{\D_\phi-1}{}_2F_1\left[\begin{matrix}\D_\phi,\D_\phi\\ 2\D_\phi\end{matrix};1-y\right]\,.
\ee
where $y=\frac{u x(1-x)}{1-x(1-v)}$. For $\D_\phi=1-\e/2$ in $d=4-\e$ dimensions, the leading term from the contact diagram is,
\be
A_c^\text{L}(z,\bz)=-\int_0^1 \frac{dx}{(1-x z)(1-x\bz)}\log \frac{z \bz x(1-x)}{1-x\bz}\,.
\ee
The discontinuity of the above contact piece is same as the leading contact Feynman diagram given by,
\be
{\rm Disc}[A_c^\text{L}(z,\bz)]=\frac{\log(\bz/z)}{\bz}\,.
\ee
So far this contact diagram does not yield anything we already do not have. To analyze further, we can consider the contact diagram for $L=2$ which corresponds to $(\nabla\phi)^4$ operator exchange. This involves,
\begin{align}
\begin{split}
A_c^\text{NL}(z,\bz)=v^{-\D_\phi}\bigg(a_{00}A_c^\text{L}(z,\bz)+\int ds dt v^s u^t \G(-t)^2\G(s+t)^2\G(\D_\phi-s)^2 (&a_{10}[s(s+t-\D_\phi)(t+\D_\phi)]\\
&+a_{01}[t(s+t)+s(s-\D_\phi)])\bigg)\,.
\end{split}
\end{align}
The $a_{00}$ term gives the usual ${\rm Disc}$ tree level Feynman Diagram. For the other two contributions, we will put $\D_\phi=1$ to get to the leading contribution. Thus to the leading order in $z\rightarrow0\,,\bz\rightarrow1$ we get, 
\begin{align}\label{NL}
\begin{split}
{\rm Disc}A_c^{\text{NL}}(z,\bz)=-&\frac{\log(1-\bz)}{\bz}\bigg(a_{00}B_0+\frac{1}{\bz^2}((a_{01}+a_{10})+(3a_{01}+5a_{10}\\
&+2(a_{01}+a_{10})B_0)(1-\bz)+2(a_{01}+a_{10})(1-\bz)^2)\bigg)\,.
\end{split}
\end{align}
This shows that we are free to include higher (and theoretically infinite) number of higher derivative contact terms, but these inclusions contribute to a constant piece to the ${\rm Disc} A(z,\bz)$ part to the leading order. In order to distinguish them, it then becomes necessary to analyze the sub leading orders as well. As a further results, we consider the case for $m+n=2$ (the next higher derivative contact term), for which,
\begin{align}\label{NNL}
\begin{split}
{\rm Disc}A_c^{\text{NNL}}(z,\bz)=-\frac{\log(1-\bz)}{\bz}&\frac{4}{\bz^4}\bigg[((4a_{20}+5a_{11}+6a_{02}+(a_{02}+a_{11}+a_{20})B_0)(1-\bz)\\
&+8(11a_{02}+9a_{11}+7a_{20}+2(a_{02}+a_{11}+a_{20})B_0)(1-\bz)^2\\
&+4(7a_{02}+8a_{11}+9a_{20}+(a_{02}+a_{11}+a_{20})B_0)(1-\bz)^3\bigg]\,.
\end{split}
\end{align}
Note that $O(1-\bz)$ and $O(1-\bz)^2$ terms in \eqref{NNL} will mix with terms at the same order in \eqref{NL}. This means, that we will have to resort to solving the equations order-by-order in $1-\bz$ recursively in the limit $1-\bz\ll z\ll 1$. In general we can write, 
\be
{\rm Disc}A_c(z,\bz)=-\frac{\log(1-\bz)}{\bz} \sum_{m,n\geq0} \frac{(1-\bz)^m}{\bz^n}\left(c_{m,n}+d_{m,n}B_0\right)\,,
\ee
where the coefficients $c_{m,n}$ and $d_{m,n}$ will encode the information about the coupling for the higher derivative contact terms. Now we add this to the usual ${\rm Disc}$ part and fix the coefficients order-by-order around $\bz\rightarrow1$. In this limit we can also approximate $B_0\approx \log z$ without loss of generality. Note that we have set $\D_\phi=1$ from the onset which only provides a single discontinuity. In general there are higher order discontinuities obtained from expanding generic $v^{\D_\phi}$ for $\D_\phi=1-\e/2+\g_\phi g^2+\dots$. This provides much more freedom to incorporate the contact terms specially to higher order in the constraint equations. 

Our basis encodes the kinematic behavior of the conformal block expansion upto some constants. Using the above form of contact terms, it is possible to suitably adjust the coefficients (the coupling $a_{mn}$) so that these constants vanish. Since there are more freedom (provided we allow for higher derivative contact terms) in these arbitrary coefficients, we can always choose these couplings in the form to get rid of additional constants. For example, the $m=0$ equation,
\be
{\rm Disc}A_c(z,\bz)=-\frac{\log(1-\bz)}{\bz} \sum_{n\geq0} \frac{c_{0,n}+d_{0,n}B_0}{\bz^n}\,,
\ee
can be added to the existing basis to adjust the additional constants proportional to $\zeta_n$. However for a better analysis, we need to keep perturbative $\D_\phi$ and further expand in $\D_\phi$ which generates higher order terms as well as $\log u$ terms to completely fix the coefficients.

\section{Discussion}\label{discussion}
We will discuss point-wise the features we have explored in the work. We will also try to elaborate on the questions we have yet to address.
\begin{itemize}
\item \cite{Alday:2017zzv} analyzed the expansion of four point function of scalars in $\e-$expansion. We explore the possibility that the basis of functions can be derived from a simple generating function systematically. The generating function works well for pure $g$ terms and mixed terms in the expansion upto $O(g^4)$. From $O(g^5)$ onwards a new class of generating functions is additionally required, albeit the number of such class of functions should be finite. 
\item The motivation of the generating function comes from diagrammatic perturbation theory for massless $\phi^4$ theory. \eqref{genreg} is a generalization of the master integrals for a large subset of loop feynman diagrams. Further, \eqref{genreg} can be deployed to rewrite both the expansion of the conformal correlator and the loop feynman diagrams in terms of the same basis of functions \footnote{For the diagrammatic computation, any regularization scheme does not alter the structural properties we are concerned with. Hence, the same building blocks used to rewrite these diagrams.}.
\item We also discuss the higher derivative contact terms within our scenario. From section \eqref{contactt}, it becomes clear that the higher derivative contact terms arrange themselves in in terms of $\bz^{-n}$. In terms of the Polyakov-Mellin bootstrap discussed in \cite{Gopakumar:2016cpb,Gopakumar:2016wkt,Gopakumar:2018xqi}, the contact terms can be used to fix the anomalous dimensions and OPE coefficients for $O(>\epsilon^3)$. In our context, the contact terms can be used to fix the constant pieces appearing in the basis expansion (\ref{mainresult}). 
\item Our technique can also be applied to other situations with $\epsilon$-expansion like boundary CFTs \cite{bissibcft}. The boundary CFT case is much simpler from the conformal correlator expansion since two-point functions are non-trivial \cite{guhabcft1,marcobcft} and have conformal blocks associated to them. In addition the crossing equation can be satisfied in boundary CFT with finite number of terms at lower orders in $\epsilon$-expansion. 
\item Another question directly connected with the developments in \cite{Mastrolia:2018uzb, Frellesvig:2019kgj, Frellesvig:2019uqt}, is to write the basis of the conformal block expansion in terms of the topologically invariant quantities like the {\it intersection numbers}. This will provide the most organized basis of expansion as demonstrated in the above mentioned references. 
\end{itemize}

\section{Acknowledgements} 
The authors thank Aninda Sinha, Shiraz Minwalla, Abhijit Gadde, Subham Duttachowdhury and Katrin Becker for useful discussions and comments on the draft. The authors specially thank Claude Duhr for helping out with PolylogTools package and some integrals.  The authors would also like to thank Surya Kiran Kanumilli and Balakrishnan Nagaraj for collaboration during the early stages of the project. SG would like to thank his advisor Katrin Becker for partial funding. KS thanks TIFR, Mumbai and McGill University for hospitality during the course of the work. KS was partially supported by World Premier Research Center Initiative (WPI), MEXT Japan at Kavli IPMU, the University of Tokyo.

\appendix
\section{Important identities}\label{impid}
We list down some important integrals that we will use throughout our calculations. We first write down result of 3-external-point integral,
\begin{align}\label{3pt}
\begin{split}
I_{a,b,c}(x_1,x_2,x_3)&=\int \frac{d^4x}{((x_1-x)^2)^{a}((x_2-x)^2)^{b}((x_3-x)^2)^{c}}\\
&=\frac{\G(\frac{a+b-c}{2})\G(\frac{a+c-b}{2})\G(\frac{b+c-a}{2})}{\G(a)\G(b)\G(c)}(x_{12}^2)^{-\frac{a+b-c}{2}}(x_{13}^2)^{-\frac{a+c-b}{2}}(x_{23}^2)^{-\frac{b+c-a}{2}}\,.
\end{split}
\end{align}
for $a+b+c=d$ and the result for 4-external points $(a_1+a_2+a_3+a_4=d)$ is, 
\begin{align}\label{4pt}
\begin{split}
&I_{\{a_i\}}(x_i)=\int  d^4 x \prod_{i=1}^4 \frac{1}{((x_i-x)^2)^{a_i}}\\
=&\frac{1}{\prod_i \G(a_i)}\frac{(x_{12}^2)^\frac{a_3+a_4-a_1-a_2}{2}(x_{14}^2)^\frac{a_3+a_2-a_1-a_4}{2}(x_{24}^2)^\frac{a_1+a_4-a_3-a_2}{2}}{(x_{13}^2)^{a_3}(x_{24}^2)^{a_4}}\\
&\times\int dsdt\ \G(-s)\G(-t)\G(s+t+a_3)\G(s+t+\frac{a_2+a_3+a_4-a_1}{2})\G(\frac{a_1+a_2-a_3-a_4}{2}-s)\\
&\times\G(\frac{a_1+a_4-a_2-a_3}{2}-t)u^s v^t\,.
\end{split}
\end{align}
We also list down the conversion of an integral from a Mellin-type to an Euler type,
\begin{align} \label{euler}
\begin{split}
& \oint ds \Gamma(a_1+s)\Gamma(a_2+s)\Gamma(b_1-s)\Gamma(b_2-s)z^{-s} \\=&\Gamma(a_1+b_1)\Gamma(a_2+b_2)\int_0^1\frac{dp}{p(1-p)}p^{b_2+a_1}(1-p)^{b_1+a_2}[1-p(1-z)]^{-b_1-a_1}
\end{split}
\end{align}
\section{Conformal Blocks: Details}\label{cbdetails}
The details of the derivation of the conformal blocks (both scalar and spin exchanges) are given here. 
\subsection{Scalar Conformal Block}
We will explicitly compute the expansion of the conformal blocks in $d=4-\e$ dimensions as a specific expansion in both the coupling$-g$ and $\e$. To start with, we will consider the specific example of scalar conformal block in the $t-$channel, in the integral representation,
\be
G_{\D,0}(1-z,1-\bz)=\frac{k_{d-\D,0}}{\g_{\l_1,0}^2}\int ds dt\ \G(\l_2-s)\G(\bar{\l}_2-s)\G(-t)^2\G(s+t)^2 (z\bz)^t ((1-z)(1-\bz))^s\,.
\ee
To explain the notation,
\be
\l_2=\D/2=\l_1\,, \bar{\l}_2=(d-\D)/2\,, k_{d-\D,\ell}=\frac{1}{(d-\D-1)_\ell}\frac{\G(\D+\ell)}{\G(h-\D)}\,,\g_{x,y}=\G(x+y)\G(x-y)\,.
\ee
We have $\D=2\D_\phi+g$, $\D_\phi=(d-2)/2$ and $d=4-\e$, we can immediately see how the expansion should work. We start by projecting out the poles from the shadow part. For this, we multiply the integral representation by a phase, 
\be
p(s)=\frac{\sin\pi(\bar{\l}_2-s)}{\sin\pi(\bar{\l}_2-\l_2)}e^{i\pi s}\,,
\ee
and performing the $t-$integral by keeping only the leading term in the $z\rightarrow0$ limit, we can write,
\begin{align}
\begin{split}
G_{\D,0}(1-z,1-\bz)&=-\frac{\G(\D)}{\G(\frac{\D}{2})^4}\G(1+\D-h)\int ds\ \frac{\G(\D/2-s)\G(s)^2}{\G(1+s+\D/2-h)}(\bz-1)^s(\log z\bz+2H_{s-1})\\
&=D_\alpha \bigg[-(z\bz)^{\alpha/2}\frac{\G(\D)}{\G(\frac{\D}{2})^4}\G(1+\D-h)\int ds\ \frac{\G(\D/2-s)\G(s)\G(s+\alpha)}
{\G(1+s+\D/2-h)}(\bz-1)^s\bigg]\bigg|_{\a=0}\,,
\end{split}
\end{align}
where $D_\a\equiv 2(\g+\pd_\a)$. Now we shift, $s\rightarrow s+\D/2$ so that, 
\begin{align}
\begin{split}
G_{\D,0}(1-z,1-\bz)=&(1-\bz)^{\D/2}D_\alpha \bigg[-(z\bz)^{\alpha/2}\frac{\G(\D)(
	\D/2)_\a}{\G(\frac{\D}{2})^2}\int ds\ \frac{\G(-s)(\D/2)_s(\D/2+\a)_s}
{(1+\D-h)_s}(\bz-1)^s\bigg]\bigg|_{\a=0}\,.
\end{split}
\end{align}
Notice that,
\be
\int ds\ \frac{\G(-s)(\D/2)_s(\D/2+\a)_s}{(1+\D-h)_s}(1-\bz)^s=-{}_2F_1[\D/2,\D/2+\a,1+\D-h,1-\bz]\,.
\ee
However the integrand inside the Euler-representation of the above is not convergent itself. We use the following transformation,
\begin{align}
\begin{split}
{}_2F_1[A_1,A_2,B_1,z]=&\frac{(2B_1-A_1-A_2+1)z-B_1}{B_1(z-1)}{}_2F_1[A_1,A_2,B_1+1,z]\\
&-\frac{(B_1-A_1+1)(B_1-A_2+1)z}{B_1(B_1+1)(z-1)}{}_2F_1[A_1,A_2,B_1+2,z]\,.
\end{split}
\end{align}
to write,
\begin{align}
\begin{split}
G_{\D,0}(1-z,1-\bz)&=-\frac{(1-\bz)^{\D/2}}{\bz}D_\alpha \bigg[(z\bz)^{\alpha/2}\frac{\G(1+\D-h)\G(\D)(
	\D/2)_\a}{\G(\frac{\D}{2})^3\G(\frac{\D}{2}-h+2)}\\
&\times\int_0^1 dx\frac{x^{\D/2-1}(1-x)^{\D/2-h+1}}{(1-x(1-\bz))^{\D/2+\a}}((h-1-\D/2)\bz+x(\bz-1)(h+\a-2-\D/2)-\D/2)\bigg] \,.
\end{split}
\end{align}
We define, 
\begin{align}
\begin{split}
\mathcal{I}_1^{\D,h}(x,\a,1-\bz)&=\frac{x^{\D/2-1}(1-x)^{\D/2-h+1}}{(1-x(1-\bz))^{\D/2+\a}}\,,\\
\mathcal{I}_2^{\D,h}(x,\a,1-\bz)&=\frac{x^{\D/2}(1-x)^{\D/2-h+1}}{(1-x(1-\bz))^{\D/2+\a}}\,.
\end{split}
\end{align}
With all these results we get,
\begin{align}
\begin{split}
G_{\D,0}(1-z,1-\bz)&=\frac{(1-\bz)^{\D/2}}{\bz}D_\alpha \bigg[(z\bz)^{\alpha/2}\frac{\G(1+\D-h)\G(\D)(
	\D/2)_\a}{\G(\frac{\D}{2})^3\G(\frac{\D}{2}-h+2)}\\
&\times\bigg((\D/2-(h-1-\D/2)\bz)\int_0^1 dx\ \mathcal{I}_1^{\D,h}(x,\a,1-\bz)\\
&+(1-\bz)(h+\a-2-\D/2)\int_0^1 dx\ \mathcal{I}_2^{\D,h}(x,\a,1-\bz)\bigg)\bigg] \,.
\end{split}
\end{align}
Finally taking the derivative {\it wrt} $\a$, we can write,
\begin{align}\label{scalfin}
\begin{split}
&G_{\D,0}(1-z,1-\bz)\\
&=\frac{(1-\bz)^{\D/2}}{\bz}\frac{\G(1+\D-h)\G(\D)}{\G(\frac{\D}{2})^3\G(\frac{\D}{2}-h+2)}\bigg[(\D/2-(h-1-\D/2)\bz)\\
&\times\int_0^1 dx\ \mathcal{I}_1^{\D,h}(x,0,1-\bz)\bigg(2H_{\D/2-1}+\log\frac{z\bz}{(1-x(1-\bz))^2}\bigg)\\
&+(1-\bz)\int_0^1 dx\ \mathcal{I}_2^{\D,h}(x,0,1-\bz)\bigg(2+(h-2-\D/2)(2H_{\D/2-1}+\log\frac{z\bz}{(1-x(1-\bz))^2})\bigg)\bigg] \,.
\end{split}
\end{align}

\subsection{$\tau=2$, $\ell\geq2$ Conformal Blocks}
We will mimic the calculation of the previous section directly from the integral representation of the conformal blocks. We start with,
\be \label{spinblock}
G_{\D,\ell}(z,\bz)=\frac{k_{d-\D,\ell}}{\g_{\l_1,0}^2}v^{\l_2}\int dsdt\ \frac{\G(-s)}{\G(1+s+\l_2-\bar{\l}_2)}\G(-t)^2\G(s+\l_2+t)^2 \a_\ell(s,t)(-v)^s u^t\,,
\ee
where $u=z\bz$, $v=(1-z)(1-\bz)$, we have first removed the effect of the shadow poles by introducing a suitable phase and further shifted $s\rightarrow s+\l_2$ so that now we can only consider the poles at $s=n$ to retrieve the phsyical conformal block. $\a_\ell(s,t)$ is the Mack polynomial, given by,
\begin{align}\label{amack}
\begin{split}
\a_\ell(s,t)=&\frac{1}{(d-2)_\ell}\sum_{m+n+p+q=\ell}\frac{(-1)^{p+n}\ell!}{m!n!p!q!}(2\bar{\l}_2+\ell-1)_{\ell-q}(2\l_2+\ell-1)_n(\bar{\l}_1-q)_q^2(\l_1-m)_m^2\\
&\times(d-2+\ell+n-q)_q(d/2-1)_{\ell-q}(d/2-1+n)_p(-s)_{p+q}(-t)_n\,.
\end{split}
\end{align}
For our purposes, $\D=2\D_\phi+\ell+g^2\g_\ell$ and $d=4-\e$ and $\D_\phi=(d-2)/2+g^2\g_\phi$. Hence, 
\begin{align}\label{lambdas}
\l_2&=\D_\phi+\frac{g^2}{2}\g_\ell\,,\ \ \bar{\l}_2=\frac{d}{2}-\D_\phi-\ell-\frac{g^2}{2}\g_\ell\,,\\
\l_1&=\D_\phi+\ell+\frac{g^2}{2}\g_\ell\,, \ \ \bar{\l}_1=\frac{d}{2}-\D_\phi-\frac{g^2}{2}\g_\ell\,.
\end{align}
Since the double discontinuity will only come from the outside factor $(1-\bz)^{\l_2}$ and we are only interested in the leading and next to leading order in the computation, it suffices to ignore the $O(g^2)$ terms in \eqref{lambdas}. Thus to the desired order of computation, we can always write, 
\begin{align}\label{ls1}
\l_2&=1-\frac{\e}{2}\,,\ \ \bar{\l}_2=1-\ell\,,\ \ \l_1=1-\frac{\e}{2}+\ell\,, \ \ \bar{\l}_1=1\,,
\end{align}
where we have neglected $O(g^2)$ contributions both from $\D_\phi$ and $\D_\ell$. The overall factors associated with the normalization of the conformal block is given by,
\be
k_{d-\D,\ell}=\frac{\G(\D+1-d/2)\G(\D+\ell)}{(d-\D-1)_\ell}\,, \ \ \g_{x,0}=\G(x)^2\,.
\ee
With the values of $\lambda_{1,2}$ given in \eqref{ls1}, it is not difficult to see that \eqref{amack} undergoes fair amount of simplifications. Firstly, the amount of sum reduces since the only term that survives is $q=0$. Then, 
\begin{align}
\begin{split}
\a_\ell(s,t)=&\frac{(d/2-1)_\ell}{(d-2)_\ell}\sum_{m+n+p=\ell}\frac{(-1)^{p+n}\ell!}{m!n!p!}(1-\ell)_{\ell}(2\l_2+\ell-1)_n(\l_1-m)_m^2(d/2-1+n)_p(-s)_p(-t)_n\\
=&\frac{(d/2-1)_\ell\G(\l_1)^2}{(d-2)_\ell\G(1-\ell)}\sum_{n+p\leq\ell}\frac{(-1)^{p+n}\ell!(2\l_2+\ell-1)_n(-s)_p(-t)_n}{(\ell-n-p)!n!p!\G(d/2-1+n+p)\G(d/2-1+n)}\\
=&\frac{(d/2-1)_\ell\G(\l_1)^2}{(d-2)_\ell\G(1-\ell)}\frac{\G(d/2-1+\ell+s)}{\G(d/2-1+\ell)}\sum_{n\leq\ell}\frac{(-1)^{n}\ell!(d+\ell-3)_n(-t)_n}{n!(\ell-n)!\G(d/2-1+n)\G(d/2-1+n+s)}\\
=&\frac{(d/2-1+s)_\ell \G(\l_1)^2}{\G(d/2-1)^2\G(1-\ell)(d-2)_\ell}{}_3F_2\bigg[\begin{matrix}-\ell,\ell+d-3,-t\\ d/2-1,d/2-1+s\end{matrix};1\bigg]\,.
\end{split}
\end{align}
Finally, performing the $n-$sum, we can write, upto the desired order, a closed form expression, given by, 
\begin{align}
\begin{split}
\frac{k_{d-\D,\ell}}{\g_{\l_1}^2}\a_\ell(s,t)=&\frac{(d/2-1+s)_\ell\G(d-2+2\ell)}{(d-2)_\ell\G(d/2-1)^2\G(d/2-1+\ell)}\ {}_3F_2\left[\begin{matrix}-\ell,\ell+d-3,-t\\ d/2-1,d/2-1+s\end{matrix};1\right]\,.
\end{split}
\end{align}
Thus the conformal block for each spin can be written as (upto $O(\e^5)$)\footnote{We have included the overall factor $(d-2)_\ell/(d/2-1)_\ell$ in the definition so that it coincides with the usual conformal block},
	\begin{align}\label{ft1}
	\begin{split}
	G_{\D,\ell}(z,\bz)=\frac{\G(d-2+2\ell) v^{\l_2}}{\G(d/2-1)\G(d/2-1+\ell)^2}\int dsdt&\ \frac{\G(-s)\G(-t)^2\G(d/2-1+s+t)^2}{\G(d/2-1+s)}(-v)^s u^t\\
	&\times   {}_3F_2\left[\begin{matrix}-\ell,\ell+d-3,-t\\ d/2-1,d/2-1+s\end{matrix};1\right]\,.
	\end{split}
	\end{align}
To proceed, we decompose ${}_3F_2$ into its integral representation,
\begin{align}
\begin{split}
&{}_3F_2\left[\begin{matrix}-\ell,\ell+d-3,-t\\ d/2-1,d/2-1+s\end{matrix};1\right]\\
&=\frac{\G(d/2-1+s)}{\G(-t)\G(d/2-1+s+t)}\int_0^1 dx x^{-t-1}(1-x)^{d/2-2+s+t}\frac{\G(1+\ell)\G(d-3)}{\G(\ell+d-3)}C_\ell^{(d-3)/2}(1-2x)\,,
\end{split}
\end{align}
where $C^\l_\ell(x)$ is the Gegenbauer polynomial. Plugging this back in \eqref{hstwist2}, we can write, 
\begin{align}\label{ft2}
\begin{split}
G_{\D,\ell}(z,\bz)=&\frac{\G(d-2+2\ell) v^{\l_2}}{\G(d/2-1)\G(d/2-1+\ell)^2}\int dsdt\ \G(-s)\G(-t)\G(d/2-1+s+t)(-v)^s u^t\\
&\times  \int_0^1 dx x^{-t-1}(1-x)^{d/2-2+s+t}\frac{\G(1+\ell)\G(d-3)}{\G(\ell+d-3)}C_\ell^{(d-3)/2}(1-2x)\,.
\end{split}
\end{align}
Now, notice that for twist$-2$ higher spin conformal blocks, $\D=2\D_\phi+\ell+g^2\g_\ell$, where,  
\be
\g_\ell=-\frac{12\g_\phi^{(2)}}{\ell(\ell+1)}\,, \ \ a_\ell=\frac{\G(d/2-1+\ell)^2\G(\ell+d-3)}{\ell!\G(d/2-1)^2\G(d-3+2\ell)}\,,
\ee
upto the order of expansion we are interested in. Thus, we define the twist$-2$ (sum over) higher spin blocks as,   
\begin{align}\label{ft3}
\begin{split}
\mathcal{G}_{2}(z,\bz)&=\left(\frac{u}{v}\right)^{\D_\phi}\sum_{\ell=2}^\infty \g_\ell^2 a_\ell G_{\tau+\ell,\ell}(z,\bz)\\
&=u^{\D_\phi}v^{\l_2-\D_\phi}\int_0^1 dx x^{-t-1}(1-x)^{d/2-2+s+t}\int dsdt\ \G(-s)\G(-t)\G(d/2-1+s+t)\\
&\times (-v)^s u^t\frac{\G(d-3)}{\G(d/2-1)^3}\sum_{\ell=2}^\infty \frac{(d-3+2\ell) }{\ell^2(\ell+1)^2}C_\ell^{(d-3)/2}(1-2x)\,.
\end{split}
\end{align}
The $s,t$ integral can be done exactly, and with the substitution $u=z\bz\,, v=(1-z)(1-\bz)$, 
\begin{align}
\begin{split}
&\int dsdt\ \G(-s)\G(-t)\G(d/2-1+s+t)u^t\left(\frac{1-x}{x}\right)^t(1-x)^s(-v)^s\\
&=\G(d/2-1) \frac{x^{d/2-1}}{(z+x(1-z))^{d/2-1}(\bz+x(1-\bz))^{d/2-1}}\,.
\end{split}
\end{align}
Substituting this in \eqref{ft3}, we can write, 
\begin{align}\label{ft4}
\begin{split}
\mathcal{G}_{2}(z,\bz)=&(z\bz)^{\D_\phi}[(1-\bz)(1-z)]^{\l_2-\D_\phi}\int_0^1 dx \frac{(x(1-x))^{d/2-2}}{(z+x(1-z))^{d/2-1}(\bz+x(1-\bz))^{d/2-1}}\\
&\times \frac{\G(d-3)}{\G(d/2-1)^2}\sum_{\ell=2}^\infty \frac{(d-3+2\ell) }{\ell^2(\ell+1)^2}C_\ell^{(d-3)/2}(1-2x)\,.
\end{split}
\end{align}
Starting from this we will extend the analysis of the scalar conformal blocks to the twist$-2$ higher spin blocks in the main text. 
\subsection{Functions in (\ref{ddiscscal})}\label{basisel}
We will write down the basis functions at each order in $\a$ starting from the leading term for $\a=0$. For $\a=0$, {\it i.e.} the leading order, we have,
\be
f_{0,0}=\frac{\log\bz-\log z}{\bz}\,,
\ee
while for $\a=1$, 
\be
f_{1,0}=\frac{\log\bz-\log z+{\rm Li}_2(1-\bz)-\zeta_2}{\bz}\,, \ \ f_{0,1}=\frac{(\log z-\log\bz)(\log z+2)+2\zeta_2}{2\bz}\,,
\ee
For the higher $i+j=\a\geq2$, we get more basis elements which can be obtained systematically from the HypExp package. For $\a=2$, 
\begin{align}
\begin{split}
f_{2,0}&=\frac{1}{12\bz}\bigg[6{\rm Li}_3\left(\frac{\bz-1}{\bz}\right)-6{\rm Li}_3(1-\bz)-\log^3\bz+6\zeta_3+14({\rm Li}_2(1-\bz)-\zeta_2)-3\log z\bz {\rm Li}_2(1-\bz)\\
&+3(\log\bz-\log z)\zeta_2-(\log z-\log \bz)(2+\log z\bz)\bigg]\,,\\
f_{1,1}&=-\frac{1}{12\bz}\bigg[\log^3\bz-12{\rm Li}_3(1-\bz)-6{\rm Li}_3\left(\frac{\bz-1}{\bz}\right)+12\zeta_3+12({\rm Li}_2(1-\bz)-2\zeta_2)+6\log z\bz{\rm Li}_2(1-\bz)\\
&+6\log z(\log\bz-\log z)-6\zeta_2(2\log z-\log\bz)\bigg]\,,\\
f_{0,2}&=\frac{1}{8\bz}\bigg[4\zeta_3+2\zeta_2(\log\bz-3\log z-4)+\log z(4+\log z)(\log\bz-\log z)\bigg]\,.
\end{split}
\end{align}
For the next order $\a=3$, 
\begin{align}\label{listofcontrib}
\begin{split}
f_{3,0}&=\frac{1}{72\bz}\bigg[3(\log z-\log\bz)(6+\log z\bz)-8\log^3\bz+6(1-3\log z\bz){\rm Li}_2(1-\bz)+18(\log z-\log\bz)({\rm Li}_3(1-\bz)+\zeta_3)\\
&-48{\rm Li}_3(1-\bz)+48{\rm Li}_3\left(\frac{\bz-1}{\bz}\right)+48\zeta_3+72{\rm Li}_4(1-\bz)-6(1+5\log z-3\log\bz)\zeta_2-63\zeta_4-36S_{2,2}(1-\bz)\bigg]\,,\\
f_{2,1}&=\frac{1}{144\bz}\bigg[2(11-6\log z)(\log\bz-\log z)-(1+6\log z)(\log\bz-\log z)(\log\bz+\log z)+\log^3\bz(-2+9\log z+6\log\bz)\\
&-2(1+24\log z\bz-9\log^2 z\bz){\rm Li}_2(1-\bz)-18{\rm Li}_2(1-\bz)^2+12(20+3\log z){\rm Li}_3(1-\bz)\\
&+6(2-9\log z\bz){\rm Li}_3\left(\frac{\bz-1}{\bz}\right)-216{\rm Li}_4(1-\bz)-72{\rm Li}_4\left(\frac{\bz-1}{\bz}\right)+2\zeta_2(13+108\log z+9\log z(\log z-\log\bz)\\
&-54\log\bz-18{\rm Li}_2(1-\bz))+378\zeta_4-240\zeta_3+36\zeta_3(3\log\bz-4\log z)+108S_{2,2}(1-\bz)\bigg]\,,\\
f_{1,2}&=\frac{1}{96\bz}\bigg[12\log^2z(\log\bz-\log z)+8\log^3\bz+4\log z\log^3\bz+3\log^4\bz+12{\rm Li}_2(1-\bz)^2-48(2+\log z){\rm Li}_3(1-\bz)\\
&-24(2+\log z\bz){\rm Li}_3\left(\frac{\bz-1}{\bz}\right)-24{\rm Li}_4\left(\frac{\bz-1}{\bz}\right)-12\zeta_2\log z(14+3\log z)+24\zeta_2(3+\log z)\log\bz\\
&+12{\rm Li}_2(1-\bz)(\log z\bz(4+\log z\bz)+2\zeta_2)+24\zeta_3(6+5\log z-3\log\bz)-252\zeta_4\bigg]\,,\\
f_{0,3}&=\frac{1}{48\bz}\bigg[42\zeta_4-12\zeta_3(2+2\log z-\log\bz)+6\zeta_2(2\log z(3+\log z)-(2+\log z)\log\bz)\\
&+\log^2z(6+\log z)(\log z-\log\bz)\bigg]\,.
\end{split}
\end{align}
\subsection{List of Integrals}\label{integrals}
The basic list of integrals contain the following kinds,
\begin{align}
\begin{split}
I_{m,n,p}(u)&=u\int_0^1 dx\frac{\log^m x\log^n(1-x)\log^p(1-u x)}{(1-u x)}\,,
\end{split}
\end{align}
The above form entail most of the integrals to be performed for the twist$-2$ integrals. The general form of the integral introduced in section \ref{twist2}, 
\be
\mathcal{I}_{m,n,p}(z,\bz)=\int_0^1 dx \frac{f(x)}{(z+x(1-z))(\bz+x(1-\bz))}=\frac{1}{\bz-z}\bigg[u_{\bz}I_{m,n,p}(u_{\bz})-u_{z}I_{m,n,p}(u_{z})\bigg]\,,
\ee
for $u_x=(x-1)/x$ and $f(x)$ has the general form,
\be
f(x)=\log^m x\log^n(1-x)\log^p(1-u x)\,,
\ee
Using this, some of the integrals used in the main text are,
\begin{align}
\mi_{0,0,0}(z,\bz)&=\frac{\log(\bz/z)}{z-\bz}\,,\ \mi_{0,1,0}(z,\bz)=\frac{{\rm Li}_2(1-\bz)-{\rm Li}_2(1-z)}{z-\bz}\,,\\
\mi_{1,0,0}(z,\bz)&=\frac{2{\rm Li}_2(1-\bz)+\log^2\bz-2{\rm Li}_2(1-z)-\log^2 z}{2(z-\bz)}\,,\\
\mi_{1,1,0}(z,\bz)&=\frac{\log(1-z)\log^2 z-\log(1-\bz)\log^2\bz+4{\rm Li}_3(1-z)+2{\rm Li}_3(z)-4{\rm Li}_3(1-\bz)-2{\rm Li}_3(\bz)}{2(z-\bz)}\,,\\
\mi_{0,3,0}(z,\bz)&=\frac{6({\rm Li}_4(1-\bz)-{\rm Li}_4(1-z))}{z-\bz}\,.
\end{align}
The other two integrals are (we only give their expressions for $z\rightarrow0$),
\begin{align}
\begin{split}
\lim_{z\rightarrow0}\mi_{2,1,0}(z,\bz)=&\frac{1}{12\bz}\bigg[\log^4\bz+24\log\bz({\rm Li}_3(1-\bz)-\zeta_3)+12\zeta_2(2{\rm Li}_2(1-\bz)+\log^2\bz)\\
&+24\bigg(\zeta_4+S_{2,2}(1-\bz)+{\rm Li}_4\bigg(\frac{\bz-1}{\bz}\bigg)-2{\rm Li}_4(1-\bz)\bigg)\bigg]\,,\\
\lim_{z\rightarrow0}\mi_{1,2,0}(z,\bz)=&\frac{1}{2\bz}\bigg[4\zeta_2{\rm Li}_2(1-\bz)+4\log\bz({\rm Li}_3(1-\bz)-\zeta_3)+4(S_{2,2}(1-\bz)-3{\rm Li}_4(1-\bz))+\zeta_4\bigg]\,.
\end{split}
\end{align}
The additional integrals are,
\begin{align}
\begin{split}
\lim_{z\rightarrow0}\mathcal{J}_1&=\lim_{z\rightarrow0}\int_0^1 dx\frac{\log((z+x(1-z))(\bz+x(1-\bz)))}{(z+x(1-z))(\bz+x(1-\bz))}\\
&=\frac{4{\rm Li}_2(1-\bz)+(\log\bz-\log z)(\log z+3\log\bz)}{2\bz}\,.
\end{split}
\end{align}
Two additional integrals we require, are
\begin{align}
\begin{split}
\mathcal{J}_2&=\int_0^1 dx\frac{\log x\log(1-x)\log((z+x(1-z))(\bz+x(1-\bz)))}{(z+x(1-z))(\bz+x(1-\bz))}\,, \\ 
\mathcal{J}_3&=\int_0^1dx\frac{\log\frac{1-x}{x}{\rm Li}_2(\frac{x}{x-1})+2{\rm Li}_3(\frac{x}{x-1})}{(z+x(1-z))(\bz+x(1-\bz))}
\end{split}
\end{align}
where $u_x=(x-1)/x$. In a similar fashin we will evaluate the final forms of the integrals $\mathcal{J}_2$ and $\mathcal{J}_3$ after the $z\rightarrow0$ limit. To evaluate $\mathcal{J}_2$ and $\mathcal{J}_3$, we use the PolyLogTools Mathematica package (\cite{polylogtool})\footnote{We thank Claude Duhr for helping us out with the integrals.}. The result of $\mathcal{J}_3$ integral is,
\begin{align}
\begin{split}
\lim_{z\rightarrow0}\mathcal{J}_3=&\frac{1}{24 \bar{z}}\bigg[-72 \text{Li}_4(1-\bar{z})-72 \text{Li}_4\left(\frac{\bar{z}-1}{\bar{z}}\right)+12 \text{Li}_2(1-\bar{z}) \log ^2\bz+24 \text{Li}_3(1-\bar{z}) \log\bz\\
&-48 \text{Li}_3\left(\frac{\bar{z}-1}{\bar{z}}\right) \log\bz+72 S_{2,2}(1-\bar{z})-24 \zeta_3 \log\bz+5 \log ^4\bz-72\zeta_4\bigg]\,.
\end{split}
\end{align}
This result has no discontinuities. The final result for the $\mathcal{J}_2$ integral is,
\begin{align}
\begin{split}
\lim_{z\rightarrow0}\mathcal{J}_2=&-\frac{1}{6\bar{z}}\Big(3 \text{Li}_2(1-\bar{z}){}^2+18\zeta_2 \text{Li}_2(1-\bar{z})-24 \text{Li}_4(1-\bar{z})+24\text{Li}_4\left(\frac{\bar{z}-1}{\bar{z}}\right)\\
&+18 \text{Li}_3(1-\bar{z}) \log \bz+6S_{2,2}(1-\bar{z})-18 \zeta_3 \log\bz+\log ^4\bz+12\zeta_2\log ^2\bz+12 \zeta_4\Big)\,.
\end{split}
\end{align}
This integral was performed using PolyLogTools and simplification was made using the rules listed in\cite{grules}. Again we find that there are no discontinuities in the final result.

\section{Perturbative Diagrams}\label{fddetails}
We will calculate perturbative diagrams up-to 3-loops in this section. Our focus is only the ring diagrams, which correspond to pure $g$ terms in the conformal block expansion. 
\subsection{Master Integral}\label{masterintegsec}
Before we begin computing loop integrals we will calculate a \textit{master} integral. This is important because this will appear in each loop calculation. The integral is given as (in t-channel form),
\begin{equation}
\int \frac{d^4x_6}{ x_{16}^{2+2g_1\delta} \, x_{26}^{2+2g_2\delta} \, x_{36}^{2+2g_3\delta} \, x_{46}^{2-2(g_1+g_2+g_3)\delta}} \,.
\end{equation}
We evaluate this integral using (\ref{4pt}) to obtain,
\begin{align}\label{genregexp}
\begin{split}
&\int \frac{d^4x_6}{ x_{16}^{2+2g_1\delta} \, x_{26}^{2+2g_2\delta} \, x_{36}^{2+2g_3\delta} \, x_{46}^{2-2(g_1+g_2+g_3)\delta}} = \\
&\int_0^1 \, dp \,\frac{p^{\delta  g_3} \Gamma ((-g_1-g_2) \delta ) ((1-p) p)^{\delta  (g_1+g_2)} (1-p
	\bar{z})^{\delta  (-g_2)-1} (1-p)^{\delta  (-g_1-g_2-g_3)} (z \bar{z})^{\delta 
		(g_1+g_2)}}{\Gamma (g_3\delta +1) \Gamma (1-(g_1+g_2+g_3) \delta )} \\
&+\int_0^1 \, dp \, \frac{\Gamma (1-g_1 \delta ) \Gamma (1-g_2 \delta ) p^{\delta  g_3} \Gamma ((g_1+g_2)
	\delta ) (1-p)^{\delta  (-g_1-g_2-g_3)} (1-p \bar{z})^{\delta  (g_1+g_2)+\delta 
		(-g_2)-1}}{\Gamma (g_1 \delta +1) \Gamma (g_2 \delta +1) \Gamma (g_3 \delta +1) \Gamma
	(1-(g_1+g_2+g_3) \delta )}\,.
\end{split}
\end{align}
It is quite cumbersome to carry around all these factors and hence we will just stick to a particular regularization scheme, $g_1=g_3$ and $g_1=-g_2$. This is the same scheme used in the main text to obtain pure $g$ terms. With this choice of regularization the \textit{master} integral becomes,
\begin{equation}\label{masterintform}
\int \frac{d^4x_6}{ x_{16}^{2-\delta} \, x_{46}^{2-\delta} \, x_{26}^{2+\delta} \, x_{36}^{2+\delta}} \,,
\end{equation}
and (\ref{genregexp}) becomes,
\begin{align}
\begin{split}
&\frac{1}{\Gamma^2(1+\delta/2)\Gamma^2(1-\delta/2)}\left(\frac{x_{14}^2}{x_{24}^2}\right)^{\delta}\frac{1}{(x_{13}^2)^{1+\delta/2}(x_{24}^2)^{1-\delta/2}}\\
&\qquad \times \int dsdt\Gamma^2(-s)\Gamma(-t)\Gamma(-t-\delta)\Gamma^2(s+t+1+\delta/2)(z\bar{z})^s(1-\bar{z})^t \,.
\end{split}
\end{align}

We need to evaluate the integral in $z\rightarrow 0$ limit. The $s=0$ residue will give us the leading $z$ piece. Just like the conformal block case, we will write the $s=0$ residue in this form,
\begin{equation}
\frac{(x_{14}^2)^\delta}{x_{13}^{2+\delta}x_{24}^{2+\delta}}\mathcal{D}_{\alpha}\int \, dt\,\frac{(1-\bar{z})^t \Gamma (-t) \Gamma (-t-\delta ) \Gamma \left(t+\frac{\delta }{2}+1\right) (z \bar{z})^{\alpha /2} \Gamma \left(t+\alpha +\frac{\delta }{2}+1\right)}{\Gamma \left(1-\frac{\delta }{2}\right)^2 \Gamma \left(\frac{\delta }{2}+1\right)^2}\,,
\end{equation}
where,
\begin{equation}
\mathcal{D}_{\alpha}=(2\gamma+2\partial_{\alpha})|_{\alpha=0}\,.
\end{equation}
The remaining t-integral can be performed using (\ref{euler}), which is then acted on by the $ \mathcal{D}_{\alpha}$ operator yields the final result,
\begin{align} \label{masterinteg}
\begin{split}
&\int \frac{d^4x_6}{ x_{16}^{2-\delta} \, x_{46}^{2-\delta} \, x_{26}^{2+\delta} \, x_{36}^{2+\delta}} = \\
\frac{(x_{14}^2)^\delta}{x_{13}^{2+\delta}x_{24}^{2+\delta}}&\int_0^1 \,dp\, \frac{(1-p)^{\delta /2} p^{-\delta /2} (1-p \bar{z})^{\frac{\delta }{2}-1} \left(2 \psi ^{(0)}\left(\frac{\delta }{2}+1\right)+2 \log (p)+\log (z \bar{z})+2 \gamma \right)}{\Gamma \left(1-\frac{\delta }{2}\right) \Gamma \left(\frac{\delta }{2}+1\right)}\,.
\end{split}
\end{align}
\subsection{Regularization prescription}\label{regp}
The perturbative diagrams come with multiple divergences. The maximum divergence corresponds to the number of loops eg a 2-loop diagram would have a quadratic divergence. We notice a similarity between the finite contribution of these diagrams and the conformal correlator expansion. With this in mind we will ignore the divergent contribution. Since finite pieces are regularization dependent we need to fix a scheme for regularization in position space. The similarity of finite piece with conformal correlator expansion prompted us to consider them as basis elements. As an example we will take the one-loop diagram whose integral form is,
\begin{equation}
I_{1L} = \int \frac{d^4x_5\,d^4x_6}{x_{15}^2 \, x_{45}^2 \, x_{56}^4 \, x_{26}^2 \, x_{36}^2} \,.
\end{equation}
The integral is divergent so we first regularize the terms with the following rules,
\begin{enumerate}
	\item Dress the propagator terms of each integrand variable with $\delta$ such that the sum of $\delta$s is zero. The sum of $\delta$s should vanish for each integrand to keep the integral conformal. 
	\item Multiply the integral with a pre-factor to cancel the $\delta$-dependence of the external points. The whole dressed-integral is now scale-invariant.
\end{enumerate}
With the above rules we obtain the following normalization,
\begin{align}\label{1L}
\begin{split}
I_{1L} =(x_{14}x_{23})^{\delta} \int \frac{d^4x_5\,d^4x_6}{x_{15}^{2+\d} \, x_{45}^{2+\d} \, x_{56}^{4-2\d} \, x_{26}^{2+\delta} \, x_{36}^{2+\delta}} \,.
\end{split}
\end{align}
In the above expression we will first perform the $x_5$ integral . The sum of $\delta$ vanishes for $x_5$ and $x_6$-integrals. Once we perform the $x_5$-integral it is still required that the $x_6-\delta$ sum vanishes. With this regularization one can perform the integral using (\ref{3pt}) and (\ref{4pt}) and then expand in $\delta$. We will neglect the divergent piece in $\frac{1}{\delta}$ and keep only the finite piece. As expected we will find that divergent pieces at higher order contain finite piece result of lower orders. Schematically we can write the three-loop results as,
\begin{equation}
=\frac{1}{\delta^3}\text(tree)+ \frac{1}{\delta^2}\text{(1-loop)}+ \frac{1}{\delta}\text{(2-loop)}+ \text{finite}
\end{equation}
\subsubsection{Generic Regularization}\label{regp2}
Let us briefly mention what happens when we take generic regularizations. Let us again consider the 1-loop integral,
\begin{align}\label{1Lgeneral}
\begin{split}
I_{1L} = \int \frac{d^4x_5\,d^4x_6}{x_{15}^{2+a\d} \, x_{45}^{2+b\d} \, x_{56}^{4+c\d} \, x_{26}^{2+d\delta} \, x_{36}^{2+e\delta}} \,.
\end{split}
\end{align}
Now with the first condition on our regularization procedure we obtain two equations,
\begin{equation}
a+b+c=0 \qquad c+d+e=0\,.
\end{equation}
Starting with 5 unknowns we have reduced our search space to 3 unknown. For rings diagrams we always have 3 unknown parameters for any loop. The was the motivation for us to construct a generating function (\ref{genfunc}) with three parameters. Starting from 3-loop we encounter additional generating function which has more parameters. With the regularization procedure under control we can start computing the loops.
\subsection{Tree Level Integral}
\begin{figure}\label{01figure}
	\centering
	\begin{tikzpicture}[scale=0.75, transform shape]
	\begin{scope}
	\node (v2) at (-3,3.5) {};
	\node (v1) at (2,-1) {};
	\node (v4) at (-3,-1) {};
	\node (v3) at (2,3.5) {};
	\draw  (v1) edge (v2);
	\draw  (v3) edge (v4);
	\node at (-3,4) {$x_3$};
	\node at (2,4) {$x_4$};
	\node at (-3,-1.5) {$x_1$};
	\node at (2,-1.5) {$x_2$};
	\node at (-0.5,2) {$x_5$};
	\end{scope}
	
	\begin{scope}[xshift=8 cm]
	\draw  (-3.5,3) edge (-1.5,1);
	\draw  (-3.5,-1) edge (-1.5,1);
	\draw  (3,-1) edge (1,1);
	\draw  (3,3) edge (1,1);
	
	\draw  (-0.25,1.0) ellipse (1.25 and 0.5);
	
	\node at (-3.5,3.5) {$x_3$};
	\node at (-3.5,-1.5) {$x_1$};
	\node at (3,3.5) {$x_4$};
	\node at (3,-1.5) {$x_2$};
	\node at (-1.5,1.5) {$x_5$};
	\node at (1,1.5) {$x_6$};
	\end{scope}
	\end{tikzpicture}
	\caption{Tree diagram and one-loop diagram}
\end{figure}
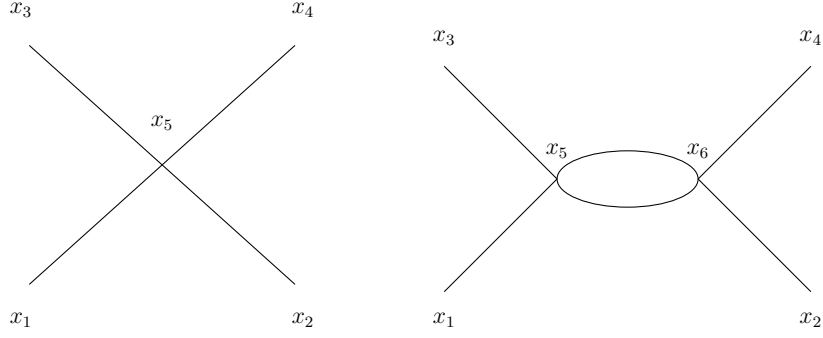

As we can see from the figure (\ref{01figure}), the tree level integral is just (\ref{masterintform}) with $\delta=0$. 
\begin{equation}
I_{0L} = \int \frac{d^4x_6}{ x_{16}^{2} \, x_{46}^{2} \, x_{36}^{2} \, x_{26}^{2}} \,.
\end{equation}
So now we plug in (\ref{masterinteg}) $\delta=0$ and obtain,
\begin{equation}
\int_0^1 \, dp\, \frac{2 \log (p)+\log (z \bar{z})}{1-p \bar{z}}\,.
\end{equation}
This integrates to ,
\begin{equation}
\frac{6 \text{Li}_2(1-\bar{z})-3 \log (z) \log (1-\bar{z})+3 \log (\bar{z}) \log (1-\bar{z})-\pi ^2}{3 \bar{z}}\,.
\end{equation}
Writing only the discontinuity we obtain,
\begin{equation}\label{treeresult}
\log(1-\bar{z})\frac{\log(z)-\log(\bar{z})}{\bar{z}}\,.
\end{equation}
\subsection{One loop ring diagram}

The one loop diagram integral is,
\begin{equation}
I_{1L} = \int \frac{d^4x_5\,d^4x_6}{x_{15}^2 \, x_{45}^2 \, x_{56}^4 \, x_{26}^2 \, x_{36}^2} \,.
\end{equation}
We will use the follow normalization prescription,
\begin{align}\label{1L}
\begin{split}
I_{1L} =(x_{14}x_{23})^{\delta} \int \frac{d^4x_5\,d^4x_6}{x_{15}^{2+\d} \, x_{45}^{2+\d} \, x_{56}^{4-2\d} \, x_{26}^{2+\delta} \, x_{36}^{2+\delta}} \,.
\end{split}
\end{align}
The justification to the prescription has already been given in (\ref{regp}). 
\newline
We first perform the $x_5$-integral (ignoring the prefactors for a moment) using (\ref{3pt}) to obtain,
\begin{align}
\begin{split}
I_{1L}= \frac{\Gamma \left(1-\frac{\delta }{2}\right)^2 \Gamma (\delta )}{\Gamma (2-\delta ) \Gamma \left(\frac{\delta }{2}+1\right)^2} \frac{1}{x_{14}^{2\delta}}   
\int \frac{d^4x_6}{ x_{16}^{2-\delta} \, x_{46}^{2-\delta} \, x_{26}^{2+\delta} \, x_{46}^{2+\delta}} \,.
\end{split}
\end{align}
The final integral is of the form of master integral and we find,
\begin{align}\label{prefinaloneloop}
\begin{split}
&I_{1L}= \frac{1}{x_{13}^{2}x_{24}^{2}}\frac{\Gamma \left(1-\frac{\delta }{2}\right)^2 \Gamma (\delta )}{\Gamma (2-\delta ) \Gamma \left(\frac{\delta }{2}+1\right)^2}  \frac{1}{x_{13}^{\delta}x_{24}^{\delta}} \\
& \times \int_0^1 \,dp\, \frac{(1-p)^{\delta /2} p^{-\delta /2} (1-p \bar{z})^{\frac{\delta }{2}-1} \left(2 \psi ^{(0)}\left(\frac{\delta }{2}+1\right)+2 \log (p)+\log (z \bar{z})+2 \gamma \right)}{\Gamma \left(1-\frac{\delta }{2}\right) \Gamma \left(\frac{\delta }{2}+1\right)}
\end{split}
\end{align}

Now we expand each term in $\delta$ and perform the integral. The divergent contribution is,
\begin{align}
\begin{split}
\frac{1}{\delta}\int_0^1\, dp \frac{2 \log (p)+\log (z \bar{z})}{ (p \bar{z}-1)}=\frac{1}{\delta}\left( \frac{\log (\bar{z})-\log (z)}{\bar{z}} \right)\,.
\end{split}
\end{align}
The finite contribution is,
\begin{align}
\begin{split}
& \int_0^1 \,dp \frac{(-2 \log (1-p \bar{z})-2 \log (1-p)+2 \log (p)+\log (1-\bar{z})-4) (2 \log (p)+\log (z \bar{z}))-4 \zeta _2}{4 p \bar{z}-4} \\
= &\log(1-\bar{z}) \big( \frac{\log (z)-\log (\bar{z})}{4 \bar{z}} \big) +  \log^2(1-\bar{z}) \big(\frac{\zeta _2-\text{Li}_2(1-\bar{z})+2 \log (z)-2 \log (\bar{z})}{2 \bar{z}} \big)\,.
\end{split}
\end{align}

\subsection{Two and three loop ring}
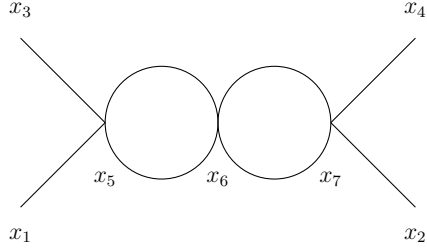
\begin{figure}
	\centering
	
	\begin{tikzpicture}[scale=0.75, transform shape]
	\begin{scope}
	\draw  plot[smooth, tension=.7] coordinates {(-3,2.5) (-1.5,1)};
	\draw  plot[smooth, tension=.7] coordinates {(-3,-0.5) (-1.5,1)};
	\draw  (-0.5,1) ellipse (1 and 1);
	\draw  (1.5,1) ellipse (1 and 1);
	\draw  plot[smooth, tension=.7] coordinates {(2.5,1) (4,2.5)};
	\draw  plot[smooth, tension=.7] coordinates {(2.5,1) (4,-0.5)};
	\node at (-3,3) {$x_3$};
	\node at (-3,-1) {$x_1$};
	\node at (-1.5,0) {$x_5$};
	\node at (0.5,0) {$x_6$};
	\node at (2.5,0) {$x_7$};
	\node at (4,3) {$x_4$};
	\node at (4,-1) {$x_2$};
	\end{scope}

	\end{tikzpicture}
	\caption{Two Loop ring}
	\label{figtwoloopring}
\end{figure}
We will write down the final results of both 2-loop and 3-loop ring diagrams here. The 2-loop integral  is given by,
\begin{equation}
I^r_{2L}=\int\frac{d^4x_5d^4x_6d^4x_7}{(x_{15}^2)(x_{45}^2)(x_{56}^2)^2(x_{67}^2)^2(x_{27}^2)(x_{37}^2)}.
\end{equation}
While the three loop integral is,
\begin{equation}
I^r_{3L}=\int\frac{d^4x_5d^4x_6d^4x_7d^4x_8}{(x_{15}^2)(x_{45}^2)(x_{56}^2)^2(x_{67}^2)^2(x_{78}^2)^2(x_{28}^2)(x_{38}^2)}.
\end{equation}

On evaluating the integral we obtain the following finite piece for 2-loop ring diagram,

\begin{align}
\begin{split}
&\log(1-\bar{z})^3\left(\frac{\log (\bar{z})-\log (z)}{24 \bar{z}}\right)
+\log(1-\bar{z})^2\left(\frac{\text{Li}_2(1-\bar{z})-\zeta _2}{4 \bar{z}}\right)\\
&-\log(1-\bar{z})\bigg(\frac{-6 \zeta _3+3 \text{Li}_2(1-\bar{z}) \log (z)+6 \text{Li}_3(1-\bar{z})-6 \text{Li}_3\left(\frac{\bar{z}-1}{\bar{z}}\right)+3 \text{Li}_2(1-\bar{z}) \log (\bar{z})}{12 \bar{z}}\\
&\frac{+3 \zeta _2 (\log (z)-\log (\bar{z}))+12 \log (z)+\log ^3(\bar{z})-12
	\log (\bar{z})}{12 \bar{z}} \bigg)  \,,
\end{split}
\end{align}

and for 3- loop diagram,

\begin{align}
\begin{split}
&\log^4(1-\bar{z})\big( \frac{\log (z)-\log (\bar{z})}{24 \bar{z}} \big)\\
&+\log^3(1-\bar{z})\big( \frac{\zeta _2-\text{Li}_2(1-\bar{z})-\log (z)+\log (\bar{z})}{6 \bar{z}} \big) \\
&+\log^2(1-\bar{z})\big( \frac{-6 \zeta _3+12\text{Li}(1-\bar{z})+3 \text{Li}_2(1-\bar{z}) \log (z)+6 \text{Li}_3(1-\bar{z})-6 \text{Li}_3\left(\frac{\bar{z}-1}{\bar{z}}\right)+3 \text{Li}_2(1-\bar{z}) \log (\bar{z})}{12 \bar{z}}\\
&\qquad +\frac{3 \zeta _2 (\log (z)-\log (\bar{z}))+24 \log (z)+\log ^3(\bar{z})-24
	\log (\bar{z})-12\zeta_2}{24 \bar{z}} \big) \\
&-\frac{\log(1-\bar{z})}{24 \bar{z}}\Big(-21 \zeta _4+24 \text{Li}_4(1-\bar{z})-6 \text{Li}_3(1-\bar{z}) \log (\bar{z})-12 S_{2,2}(1-\bar{z})+18 \zeta _3 \log (\bar{z}) \\
& -12 \zeta _3+12 \text{Li}_3(1-\bar{z})-12 \text{Li}_3\left(\frac{\bar{z}-1}{\bar{z}}\right)+6 \text{Li}_2(1-\bar{z}) \log (\bar{z})-6 \zeta _2 \log (\bar{z})+2 \log ^3(\bar{z})\\
& + 48 \text{Li}_2(1-\bar{z})-48 \zeta _2 \\
& +\log(z) \big( 6 \zeta _2-18 \zeta _3+6 \text{Li}_2(1-\bar{z})+6 \text{Li}_3(1-\bar{z}) \big) \\
& +48\log(z) - 48 \log(\bar{z}) \Big)\,.
\end{split}
\end{align}
Let us close this appendix with a few observations. The highest order discontinuity is always the tree level result. The lower order discontinuities at a given loop can be written in terms of discontinuities appearing in a lower loop diagram. The terms which appear in the discontinuities are similar to  one that occur in conformal correlator expansion.  

\section{$Li_2(1-\bar{z})^2$ origin}\label{li2}
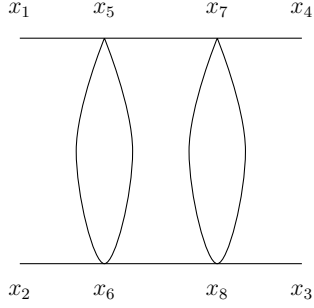
\begin{figure}
	\centering
	
	\begin{tikzpicture}[scale=0.75, transform shape]
	\draw (-5,4) -- (0,4);
	\draw (-5,0) -- (0,0);
	
	\draw  plot[smooth, tension=.7] coordinates {(-3.5,4) (-4,2) (-3.5,0) (-3,2) (-3.5,4)};
	
	\draw  plot[smooth, tension=.7] coordinates {(-1.5,4) (-2,2) (-1.5,0) (-1,2) (-1.5,4)};
	\node at (-5,4.5) {$x_1$};
	\node at (-3.5,4.5) {$x_5$};
	\node at (-1.5,4.5) {$x_7$};
	\node at (0,4.5) {$x_4$};
	
	\node at (-5,-0.5) {$x_2$};
	\node at (-3.5,-0.5) {$x_6$};
	\node at (-1.5,-0.5) {$x_8$};
	\node at (0,-0.5) {$x_3$};
	
	\end{tikzpicture}
	\caption{Three Loop ladder}
	\label{figtwoloopring}
\end{figure}
In this section we will demonstrate the origin of $\text{Li}_2^2(1-\bar{z})$ from a new generating function. We encountered these terms in the conformal block expansion at 5th order. This corresponds to three loop in the diagrammatic expansion. At three loops we encounter diagrams which contribute to an additional generating function. Ladder diagram shown in the figure above is the one that generates such a term. We will report its integral,
\begin{equation}
I=\int\frac{d^4x_5d^4x_6d^4x_7d^4x_8}{(x_{15}^2)(x_{47}^2)(x_{26}^2)(x_{38}^2)(x_{56}^2)^2(x_{78}^2)^2(x_{57}^2)(x_{68}^2)} \,.
\end{equation}
Proceeding along the lines of (\ref{regp2}) we see that the most general regularization of the above integral has 8 parameters and we have 4 equations. This tells us that there would be 4 independent parameters and this would be our new generating function.
We will regularize the integral as,
\begin{equation}
I\sim \int\frac{d^4x_5d^4x_6d^4x_7d^4x_8}{(x_{15}^2)^{1-\delta/2}(x_{47}^2)^{1+3\delta/2}(x_{26}^2)^{1-3\delta/2}(x_{38}^2)^{1+\delta/2}(x_{56}^{2})^{2+\delta}(x_{78}^2)^{2-\delta}(x_{57}^2)^{1-\delta/2}(x_{68}^2)^{1+\delta/2}} \,.
\end{equation}
Performing the $x_5$ and $x_6$ integral one obtains,
\begin{equation}
I\sim \Gamma(\delta)\Gamma(-\delta)\int \frac{d^4x_6d^4x_7}{(x_{17})^{-\delta}(x_{36})^{\delta}(x_{76})^{2}(x_{47})^{1+3\delta/2}(x_{26})^{1-3\delta/2}(x_{37})^{1-\frac{\delta}{2}}(x_{16})^{1+\frac{\delta}{2}}}
\end{equation}
Our goal here is to schematically show the piece we want and so we will use a small trick to obtain the term of interest. For the lower loop calculations we first performed the integral and then took the $\delta \rightarrow 0$ limit. However this process should commute. In that spirit we will take $\delta \rightarrow 0$ limit right now and focus on a particular term. The term that we want to focus on is the above equation with $(x_{17})^{-\delta}$ set to 1. This is a geniuine term which will appear when one takes the $\delta \rightarrow 0$ limit. Focusing on this terms is sufficient to generate the discontinuity we want.
\begin{equation}
I\sim \Gamma(\delta)\Gamma(-\delta)\int \frac{d^4x_6d^4x_7}{(x_{36})^{\delta}(x_{76})^{2}(x_{47})^{1+3\delta/2}(x_{26})^{1-3\delta/2}(x_{37})^{1-\frac{\delta}{2}}(x_{16})^{1+\frac{\delta}{2}}}.
\end{equation}
We can now perform the $x_7$ integral which gives rise to,
\begin{equation}
I\sim \Gamma(\delta)^2\Gamma(-\delta)^2 \int_0^1 \,dp\, \frac{(1-p)^{-\delta /2} p^{3\delta /2} (1-p \bar{z})^{-\frac{\delta }{2}}}{1-p\bar{z}}\,.
\end{equation}
The integrand needs to be expanded to a maximum of fourth order in $\delta$ to perform the integral. It turns out that the fourth order term gives rise to $Li_2(1-\bar{z})^2$
\bibliographystyle{JHEP}
\bibliography{genfunc}

\end{document}